\documentclass[%
preprint,
preprintnumbers,
amsmath,amssymb,
aps, showkeys
]{revtex4-2}

\usepackage{graphicx}% Include figure files
\usepackage{dcolumn}% Align table columns on decimal point
\usepackage{bm}% bold math
\usepackage[colorlinks=true, citecolor=blue, linkcolor=blue, urlcolor=blue]{hyperref}

\usepackage[separate-uncertainty=true]{siunitx}
\DeclareSIUnit\angstrom{\textup{~\AA}}

\usepackage{ulem}
\usepackage{mathrsfs}
\usepackage{upgreek}
\usepackage{amsmath}
\usepackage{graphicx}
\usepackage{nicefrac}
\usepackage{calligra}
\usepackage[version=4]{mhchem}
\usepackage{xspace}
\usepackage[percent]{overpic} 
\usepackage{enumitem} 
\newcommand{\xray}{X-ray\xspace}
\newcommand{\xrays}{X-rays\xspace}
\newcommand{\ie}{\textit{i.\,e.}\xspace}
\newcommand{\eg}{\textit{e.\,g.}\xspace}
\newcommand{\etal}{\textit{et\,al.}\xspace}
\renewcommand{\P}[1]{\mathcal{P}^{#1}}
\newcommand{\A}[1]{\mathcal{A}^{#1}}
\newcommand{\isoG}[2]{\mathcal{G}(#1, #2)}
\newcommand{\isoF}[2]{\mathcal{F}(#1, #2)}
\newcommand{\isog}{\mathcal{G}}
\newcommand{\isof}{\mathcal{F}}
\newcommand{\larsmo}{\ce{(La_{0.5}Sr_{1.5})MnO_4}\xspace}

\newcommand{\ta}[1]{\mathbf{t}_{\rm{#1}}}

\usepackage{hyperref}
\hypersetup{colorlinks=true, linkcolor=blue, citecolor=blue, urlcolor=blue}

\begin{document}

\title{A review on the Parameter Space Concept and its use for crystal structure determination}
%\thanks{Dedicated to our revered teacher Prof. Karl Fischer}%

\author{Matthias Zschornak}
\email{matthias.zschornak@physik.tu-freiberg.de}
\affiliation{Technical Physics, University of Applied Sciences, Friedrich-List-Platz 1, D-01069 Dresden, Germany\\
Center for Efficient High Temperature Processes and Materials Conversion ZeHS, TU Bergakademie Freiberg, Winklerstr.~5, D-09596 Freiberg, Germany}

\author{Muthu Vallinayagam}
\affiliation{Technical Physics, University of Applied Sciences, Friedrich-List-Platz 1, D-01069 Dresden, Germany\\
Center for Efficient High Temperature Processes and Materials Conversion ZeHS, TU Bergakademie Freiberg, Winklerstr.~5, D-09596 Freiberg, Germany }

\author{Melanie Nentwich}
\affiliation{Deutsches Elektronen-Synchrotron DESY, Notkestr.~85, D-22607 Hamburg, Germany}

\author{Dirk C. Meyer}
\affiliation{Center for Efficient High Temperature Processes and Materials Conversion ZeHS, TU Bergakademie Freiberg, Winklerstr.~5, D-09596 Freiberg, Germany\\Institute of Experimental Physics, TU Bergakademie Freiberg, Leipziger Str.~23, D-09596 Freiberg, Germany}

\author{Karl~F. Fischer$^\dag$\footnote[1]{deceased on October 29th, 2023}}
\affiliation{Institute of Experimental Physics, Universit{\"a}t des Saarlandes, D-66123 Saarbr{\"u}cken, Germany}

\collaboration{Dedicated to our revered teacher Prof. Karl Fischer}%\noaffiliation

\date{\today}

\begin{abstract}
We present a comprehensive review of the emerging crystal structure determination method Parameter Space Concept (PSC), which solves and refines either partial or complete crystal structures by mapping each experimental or theoretical observation as a geometric interpretation, bypassing the conventional Fourier inversion. The PSC utilizes only a few \xray (equivalently neutron) diffraction amplitudes or intensities and turns them into piecewise analytic hyper-surfaces, called isosurfaces, embedded in a higher-dimensional orthonormal Cartesian space (Parameter Space (PS)). It reformulates the crystal determination task into a geometric interpretation, searching for a common intersection point of different isosurfaces. The art of defining various kinds of isosurfaces, based on signs, amplitude or intensity values, normalized ratios, and ranking of reflections, offers multiple choices of adapting methods in PSC based on available experimental or theoretical observations. The elegance of PSC stems from one-dimensional (1-dim.) projections of atomic coordinates, enabling the construction of full three-dimensional crystal structures by combining multiple projections. By these means, the user may explore homometric and non-homometric solutions within both centric (well-established) and acentric (conceptually applicable) structures with spatial resolution remarkably down to \si{\pico\meter}, even with a limited number of diffraction reflections. Having demonstrated the potential of PSC for various synthetic structures and exemplarily verified direct application to realistic crystals, the origin and development of PSC methods are coherently discussed in this review. Notably, this review emphasizes the theoretical foundations, computational strategies, and potential extensions of PSC, outlining the roadmap for future applications of PSC in the broader context of structure determination from experimental diffraction observations. 
\end{abstract}


\maketitle

\section{\label{sec:intro}Introduction}
The development of PSC originates from the fundamental description of structural parameters in a high-dimensional space and the need for its precise mathematical formulation. Presenting these aspects, this section is organized into different parts to provide the general background and motivation for PSC (see Sec. \ref{sec:introgeneral}), introduce the concept of the parameter space and the role of the structure vector (see Sec.~\ref{sec:psdefinition}), outline the PSC as a framework for structure determination (see Sec. \ref{sec:solutionfinding}), and finally discuss the use of isosurfaces, which are a central tool to visualizing correlations and solutions within the PSC approach (see Sec.~\ref{sec:isosurface}).

\subsection{\label{sec:introgeneral}Motivation and General Background}
A crystallographic $m$-atomic structure, or more precisely, the scattering density $\rho(\mathbf{r})$ thereof, is parameterized in terms of the fractional coordinates of its maxima at $\mathbf{r}_j = \left(x_j,y_j,z_j\right)$, \ie atom sites $j = 1, \ldots, m;~\text{and } 0 \le x, y, z < 1$ in the unit cell of the 3-dim. periodic lattice. The density $\rho(\mathbf{r})$ can be mapped at (sub-)atomic resolution by Fourier summation (see Equ.~\eqref{eq:rho}). Thus, it is accessible for interpretation by biologists, chemists, material scientists, mineralogists, pharmacists, or physicists who recognize whole molecules or molecular fragments, their configurations, and spatial arrangements.  

The principle of structure solution with PSC is based on the inversion of a sufficiently large body of structure factor equations, modeling the relations between (i) the atomic coordinates $x_j, y_j, z_j$ and (ii) their individual atomic scattering factors $f_j$ (with $f_j$ here -- and only here -- meaning the effective $f_j$ reduced by thermal smearing) on one side and the observed structure factors $F_{\mathrm{obs}}(hkl)$ on the other, which resembles the coherent sum of all scattering contributions of all atoms $m$. For the full elementary cell housing $m$ symmetrically independent atoms (all in general positions in acentric cells or $2m$ in centric ones)

\begin{align}
    F(hkl) &= \sum_{j=1}^m f_j \cdot e^{2\pi i(hx_j+ky_j+lz_j)} && \text{or} && F(hkl) = 2 \cdot \sum_{j=1}^m f_j \cdot \cos \big(2\pi (hx_j+ky_j+lz_j)\big ) \label{eq:Facentric} \\
    F(hkl) &= A(hkl) + i \cdot B(hkl)  && \text{or} && F(hkl) = 2 \cdot A(hkl) \label{eq:Fcentric} \\
    &\text{(acentric, space group \texttt{P1})}  && \text{ } && \text{(centrosymmetric, space group \texttt{P$\bar{1}$} )} \nonumber
\end{align}

(with $f$ here -- and only here -- meaning the effective $f$ reduced by thermal smearing).

The observations $F_{\mathrm{obs}}$ are recorded from the net intensities of the reflections $(hkl)$ at the nodes of the reciprocal lattice. The inversion cannot be done analytically for structures with $m>3$. Instead, it is usually approximated by Fourier summation of the experimental structure factors $F_{\mathrm{obs}}(hkl)$ from, say, a spherical part of reciprocal space, \ie by applying an optical point-of-view and thus corresponding to an \xray microscope (magnification factor of order $\nicefrac{\lambda_{\mathrm{visible}}~}{\lambda_{\mathrm{\xrays}}}$):
\begin{equation}
    \rho(x,y,z) \approx \frac{K}{V} \cdot \sum_{hkl} F_{\mathrm{obs}} \cdot e^{-\,2\pi i(hx+ky+lz)},
    \label{eq:rho}
\end{equation}
where $V$ and $K$ are the unit cell volume and an unknown scaling factor, respectively.

The straightforward summation is, however, prevented by the lack of phase information as the scattering experiment yields only the amplitudes $K\cdot\left|F_{\mathrm{obs}}(hkl)\right|$ without phases from the reflection intensities being proportional to $\left|F_{\mathrm{obs}}(hkl)\right|^2$~\cite{WH1995}. The solution to this phase problem by the so-called Direct Methods (DM)~\cite{AM1938, HH1950, WH1995} was the decisive leap forward and was honored by the Nobel Prize for Chemistry (1985, to J. Karle and H. Hauptman). The DM are based on statistical relations between reflections whose intensities exceed expectation values (usually those of a random distribution of atoms). Thus, DM has some principally weak points (see the following paragraphs).

An older concept calculates the distribution of the interatomic vector density $\rho(\mathbf{r})\ast\rho(\mathbf{-r})$ ($\ast$ means convolution) as Fourier inversion of the experimental $\left|F_{\mathrm{obs}}(hkl)\right|^2$. It must be deconvoluted for structural interpretation, a task that becomes less and less practicable as $m$ increases (De-convolution is, however, easy for a few heavy atoms). In recent decades, crystallographic structure solution and determination methods have been developed to a degree that, in most cases, black-box work suffices to come up with correct results. 

Interested in exploring what can be achieved from amplitudes alone and without Fourier summation, the pioneers of PSC - Fischer, Kirfel, and Zimmermann - soon came across ideas of the early days (cf. Ref.~\cite{Ott27}), which could not be pursued at that time due to the missing computing facilities. This situation, which is much alleviated today, inspired them to inspect options of direct space structure determination, \ie of finding at once their $\mathbf{r}_j = \left(x_j,y_j,z_j\right)$ for all $m$. Their approach needs no reflection phase information and automatically includes deconvolution of the easily accessible interatomic vector density mentioned above, one of the weak points that are inherent to Fourier (Patterson) summation. 

Others are
\begin{enumerate}[label=(\roman*)]
    \item Accidentally vanishing or very weak reflection amplitudes do not significantly contribute to the Fourier sum and are therefore (often) routinely neglected, due to the unknown phase.
    \item Measurements must cover a (symmetrically) unique part of the spherical shell in reciprocal space to avoid systematic imaging errors (density map distortions).
    \item The resolution of the scattering density maps is limited by the maximum of $\sin{\theta}/\lambda$, \ie the largest scattering length recorded by the experiment.
    \item Finally, there is no simple and objective method to prove the uniqueness of a structure model developed.
\end{enumerate}

These shortcomings may be partially avoided by directly mapping a structure, though possibly, at the expense of some confrontations with new and unexpected problems of unknown importance, see Sec.~\ref{sec:solugeneral}.

To address the above issues, Fischer \etal proposed and developed PSC over the last decades to deal primarily with centrosymmetric structures in the initial phase. Having proposed the PSC, Fischer \etal devised the formalism for various methodologies to handle the atoms as equi-point and non-equi-point scatterers \cite{FKZ05}. They came up with approaches to specifically handle the diffraction data for PSC~\cite{KFZ06, KF09} and showed how to utilize data from different wavelengths~\cite{KF10}. Shortly, they demonstrated the potential of PSC by solving a molecule structure consisting of 11 atoms~\cite{FKZ08} and summarized the methodical development until 2009~\cite{ZF09}. Following the work of Fischer \etal, Zschornak \etal further utilized PSC to resolve the debatable La and Sr coordinates (known as split position)~\cite{ZW22}, and developed the solid algorithm for one of the PSC techniques, the linear approximation approach~\cite{VNZ2024}.

\begin{figure}
    \centering
    \includegraphics[width=1\textwidth, clip, trim=0.01in 0.01in 0.01in 0.01in]{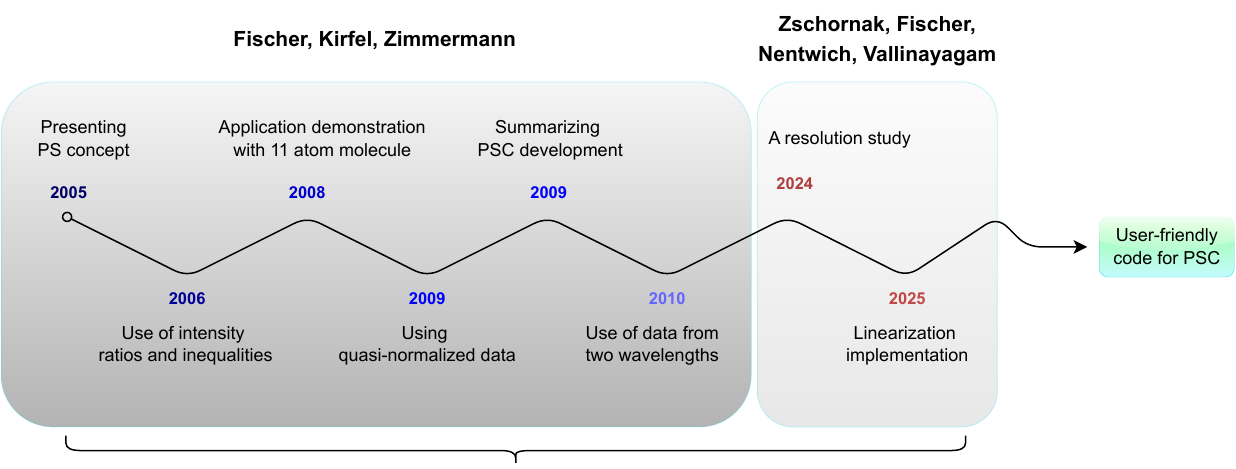}
    \caption{Development of PSC over the last decades. The final goal is to have PSC available as a complete, user-friendly simulation package.}
    \label{fgr:timeline}
\end{figure}

\subsection[Parameter space and structure vector]{\label{sec:psdefinition}Parameter space $\P{}$ and structure vector $\mathbf{t}$ }
PSC methodology looks at the inversion problem of the structure factor equation from a point-of-view different from the optical one: Assume $m$ (symmetrically) independent atoms in the ``unit cell'', for simplicity in the acentric space group \texttt{P1}, where the origin is fixed by one additional atom, or in the centric \texttt{P$\bar{1}$} with origin at an allowed site (coordinates $0$ and/or $\nicefrac{1}{2}$). Thus, neglecting all other (non-geometrical) parameters \eg, different atomic scattering powers $f_j$ and their thermal displacements affecting $f_j$, there are $3m$ independent unknown parameters. Such a model can be mapped as a vector $\mathbf{t} = \mathbf{t}_x \otimes \mathbf{t}_x \otimes \mathbf{t}_z$ with $\mathbf{t}_x(x_1,x_2,\ldots,\ x_m)$ (structure vector), or as a point in a $3m$-dim. parameter space $\P{}$ such that the true structure is given by the true structure vector $\ta{s}$. This space is orthonormal and periodic. Thus, it possesses a unit cell (including its origin) denoted by $\P{3m}$, which may be regarded as a finite subspace of the Hilbert space. This cell contains all possible geometrical structures, including $\ta{s}$, because any vector in $\P{3m}$ stands either for the structure $\ta{s}$ or for an approximate solution $\ta{a}$, or for a failure. If $\ta{a}$ is close to $\ta{s}$, $\left|\ta{s}-\ta{a}\right|$ is small and $\ta{a}$ can be expected to represent a refinable model. Dividing each dimension of the unit cell of $\P{3m}$ into $n=\nicefrac{1}{\Delta}$ equal distances fosters a grid of $n^{3m}$ test vectors $\ta{t}$. One or more of them may qualify to be a possible $\ta{a}$ as shown below in Eqs.~\eqref{eq:g_acentric-centric} (for more details see reference~\cite{FKZ05}). Hence, the objective of a structure determination is to find the \textit{best} set of $3m$ parameters $x_j,y_j,z_j; j=1,\ldots,m$. In the present context, the best $3m$ parameters represent a global minimum with best correspondance between observed reflection amplitudes/intensities and those of the model, see Sec.~\ref{sec:globalminimum} for further discussion.

A single global minimum is indicated at $\ta{s}$, if all but one possible solution vectors are rejected, \ie if only one symmetrically independent $\ta{s}$ in $\P{3m}$ signals a correct and refinable solution. Then, the structure determination is uniquely defined based on the data set used. If two or more such minima are found at vectors $\ta{s,1}$, $\ta{s,2}$, etc., the associated structures are homometric, producing indentical diffraction patterns, see Ref.~\cite{ZF09}, there Sec.~7, and Ref.~\cite{FKZ05}, there example~3. If, 
in addition, the structures are unifiable under symmetrical operations, the structures are also congruent. 

\subsection[Parameter Space Concept for finding ta ~ ts]{\label{sec:solutionfinding}Parameter Space Concept for finding approximant solutions}
We call the option to determine a structure straightforwardly in direct space the Parameter Space Concept (PSC). Parameter spaces are a well-known tool in informatics~\cite{Abreu2023}, but have not been used yet in crystal structure determination. After pursuing the underlying idea, trying various approaches as well as different numbers and kinds of observations on numerous hypothetical structures as well as on potential applications, the present paper addresses the following positive principal aspects of PSC: 
\begin{enumerate}[label=(\roman*)]
    \item Experimental data are integrated Bragg reflection intensities at the nodes of the reciprocal lattice or therefrom derived quantities like structure amplitudes $K \cdot \left|F_{\mathrm{obs}}(hkl)\right|$.
    \item Reflection phase information is not necessary and deconvolution of Patterson densities is circumvented since $\mathbf{t}_s$ yields all interatomic distances.
    \item Assumed on an absolute scale and free from statistical and/or systematic errors, $3m$ independent observations $\left|F_{\mathrm{obs}}(hkl)\right|$ suffice for a full structure determination of a crystal with $m$ atoms with the theoretically infinite (spatial) resolution of the positional parameters. Hence, a complete data set is unnecessary. 
    \item Each accidentally vanishing amplitude reduces a centro-symmetric $m$-dim. problem of unknown parameters to $m-1$ dimensions, as known since 1928~\cite{Ott27}. This reduction is $m-2$ dimension for acentric structures because both real and imaginary parts of the structure factor $F$ are zero, simultaneously. The same applies to the PSC methodology.
    
    \item\label{item:principal_aspect_v} Since any structure solution is obtained at once (as full solution vector), optimization of a first incomplete partial solution of less than $3m$ atomic coordinates (as is the case for the Patterson method) is not required.
\end{enumerate}

For a unique result see the end of Sec.~\ref{sec:psdefinition} and references~\cite{FKZ05, KFZ06, FKZ08, KF09, ZF09, KF10, KF05}. Note the varying notations in these references during the years of development. The negative principal aspects of PSC are due to the (a) approximations to the pure geometry of the structure (\eg to different atomic scattering coefficients $f_j$) and (b) quality, number, and independence of processed experimental data. They will be mentioned duly while various examples are discussed below. The practical merits and demerits of PSC will be treated together with various methods for locating $\ta{a} \approx \ta{s}$.

We shall restrict $\P{3m}$ to $\P{m}$, \ie from 3-dim. structures to 1-dim. projections, in most sections of the paper (for reasons and more details see Sec.~\ref{sec:preference}). Thus, we usually deal with the $m$ atomic fractional coordinates of a 1-dim. projection of the crystal structure onto a selected direction, \eg along $[100]$, with $m$ $x$-parameters. The observations needed for the solution are the reflection amplitudes $K \cdot \left|F_{\mathrm{obs}}(h00)\right|$ or quantities derived therefrom. Then, it is necessary to reconstruct the 3-dim. structure from a few solved 1-dim. projections, which is a more geometric task.

Up to now, three main strategies have been established for evaluating structural parameters using PSC techniques based on amplitudes or intensities. Option A (Sec. \ref{sec:optionA}) relies on calculating various kinds of figures of merit (FOM) for the grid points and identifying the global optimum. In this case, no construction of isosurfaces is required, which makes the procedure relatively straightforward. Option B (Sec.~\ref{sec:optionB}) and Option C (Sec.~\ref{sec:optionC}) are based on the concept of isosurfaces of (observed) amplitudes or intensities and their parametrization in the parameter space. Option B seeks the intersection point of all relevant isosurfaces, which, in principle, yields one or more solutions that directly correspond to the experimental data. Finally, Option C adopts a stepwise reduction approach: the parameter space is successively constrained by inequalities between carefully selected pairs of isosurfaces. This process continues until the remaining $m$-dimensional solution region becomes sufficiently small, at which point one or several solutions can be identified, for instance by determining the center of gravity of the region. While Options B and C provide deeper insight into the parameter correlations, they require numerical descriptions or approximations. In particular, Option C has so far retained a more academic character due to its computational complexity and reliance on numerical treatment. Technical details of all three options are discussed in Chapter 2.

\subsection{\label{sec:isosurface}Isosurfaces}
The fundamental requirement for structure determination is the knowledge of the Miller indices $hkl$ and their associated intensities $I(hkl)$, which are related to the structure factor $F$ as $I(hkl) \sim \left|F_{\mathrm{obs}}(hkl)\right|^2$. The structure factor for projections of the crystal structure can be expressed as~\cite{FKZ05, VNZ2024}:

\begin{subequations}
    \begin{align}
        F(h00) &= F(h) =  \sum_{j=1}^{m} f_j\,\cos(2\pi h x_j) + i\cdot\sum_{j=1}^{m} f_j\,\sin(2\pi h x_j) \quad  && \text{(acentric)}, \label{eq:F_acentric}\\
        F(h00) &= F(h) = 2\cdot \sum_{j=1}^{m} f_j\,\cos(2\pi h x_j)  \quad  && \text{(centric)}, \label{eq:F_centric}
    \end{align}
    \label{eq:F_acentric-centric}
\end{subequations}
where $m$ is the number of atoms in the unitcell and $f_j$ the respective atomic scattering factor of atom $j$. Enforcing the approximation of equivalent scattering for all atoms in the unitcell, \ie, reducing the structure to an assemblage of equal and point-like scatterers of scattering power $f=1$ (cf. Sec. 4.1 in Ref.~\cite{ZF09}), 
leads to a quantity that describes solely the geometric arrangements of the structure,
the Geometric Structure Factor~$G$
\begin{subequations}
    \begin{align}
        G(h00) &= G(h) = \sum_{j=1}^{m} \cos(2\pi h x_j) + i\cdot\sum_{j=1}^{m} \sin(2\pi h x_j) && \text{(acentric)}, \label{eq:g_acentric}\\
        G(h00) &= G(h) = 2\cdot \sum_{j=1}^{m} \cos(2\pi h x_j) && \text{(centric)}. \label{eq:g_centric}
    \end{align}
    \label{eq:g_acentric-centric}
\end{subequations}
Both $F$ as well as $G$ generally contain a complex phase (acentric) or sign (centric), next to the magnitudes $|F|$ and $|G| \equiv g$. Whereas the magnitudes are easily accessible by experimental \xray diffraction (XRD) measurements, more effort has to be made to extract the phase information, \eg by resonant methods \cite{Tobola2010, Adachito5111, Lefevreto5141, Liuto5138, Macke2016, Penanb5314, Zschornak2014}.

\begin{figure}
    \centering
    \includegraphics[width=1\textwidth,clip, trim=0.05in 0in 0in 0.in]{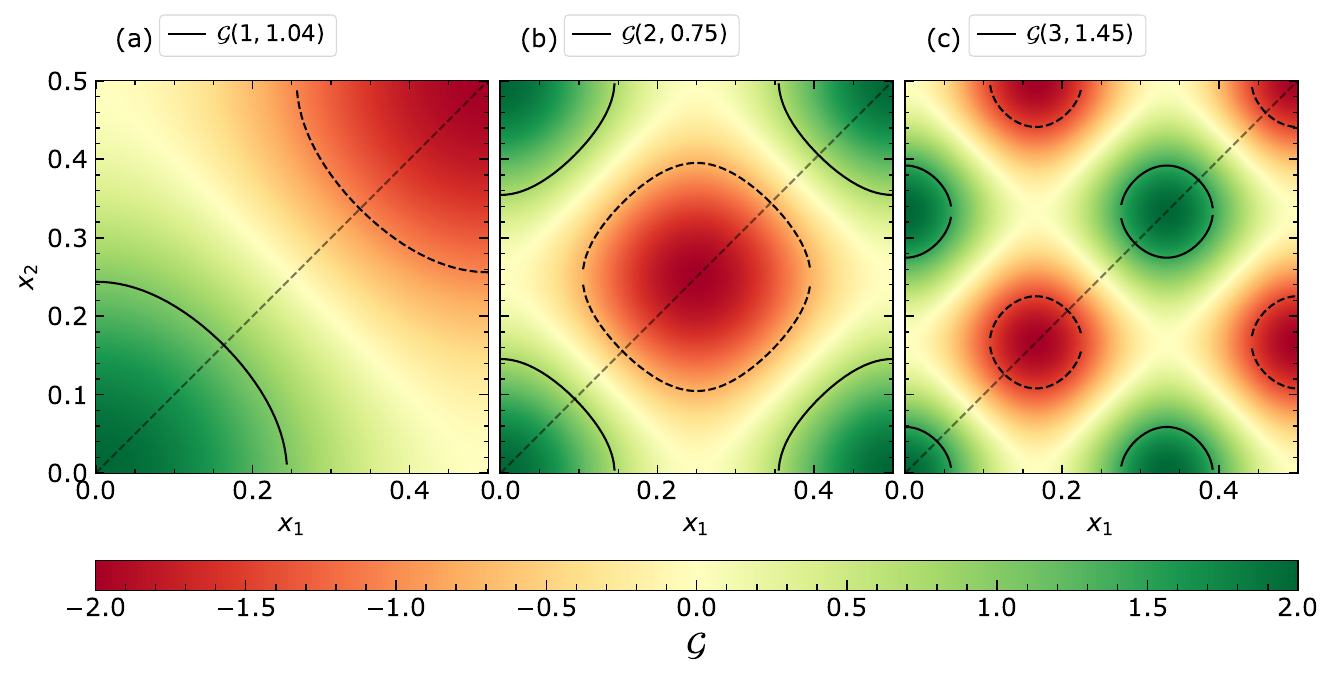}
    \caption{Example for typical isosurfaces in parameter space concept, generated for an arbitrary centrosymmetric structure consisting of two atoms. The atomic coordinates $(x_1, x_2)$ are varied in the limits $[0, 1/2]$, \ie covering half of the unitcell of the crystal structure. The isosurfaces $\isoG{h}{g(h)}$ of magnitudes $g$ equal to (a) 1.04, (b) 0.75, and (c) 1.45 for reflections $h=1$, $h=2$, and $h=3$, respectively, are highlighted by the lines. The solid and dashed lines represent the positive and negative amplitudes. The color map represents the calculated amplitude for a combination of atomic coordinates within the assumed limits. Recreated from Ref.~\cite{VNZ2024} with permission.}
    \label{fgr:isosurfaces}
\end{figure}

Each isosurface, say of projections $\left|F(hkl)\right| = \left|F(h)\right|$ or their respective squares, is geometrically represented in $\P{m}$ by a $(m-1)$-dim. manifold $\mathcal{M}$ (thereafter given in one of the quantity or observable, \eg $\mathcal{I}$ for $I$, $\isof$ for $F$, $\isog$ for $G$ etc.) \ie a subspace containing all parameter combinations $x_j$, which obey the given $\left|F(h)\right|$ (see pages 644--647 in Ref.~\cite{FKZ05} and also Ref.~\cite{ZF09}). Correspondingly, any other quantity derived from $\left|F(h)\right|$, as the quasi-normalized $e(h,n)$~\cite{KF09}, is likewise represented in $\P{m}$ as a $(m-1)$-dim. isosurface that houses, correlates, and restricts the possible parameters $x_j$ according to the value of the corresponding experimental observation, say $\left|F_\mathrm{obs}(h)\right|$. Thus, each isosurface depends on $h$ and on the value of $\left|F_\mathrm{obs}(h)\right|$ and may be designated by $\isoF{h}{|F(h)|}$ or $\isoG{h}{g(h)}$, respectively. The structure is determined when the intersections of $m$ isosurfaces of the same type ($\mathcal{I}$, $\isof$, $\isog$, etc.) are found, each common to all $n\geq m$ observations. More details can be found in Refs.~\cite{FKZ05, KFZ06, FKZ08, KF09, ZF09, KF10, VNZ2024} and in Sec.~\ref{sec:optionB}.

\begin{figure}
    \centering
    \includegraphics[width=0.9\textwidth]{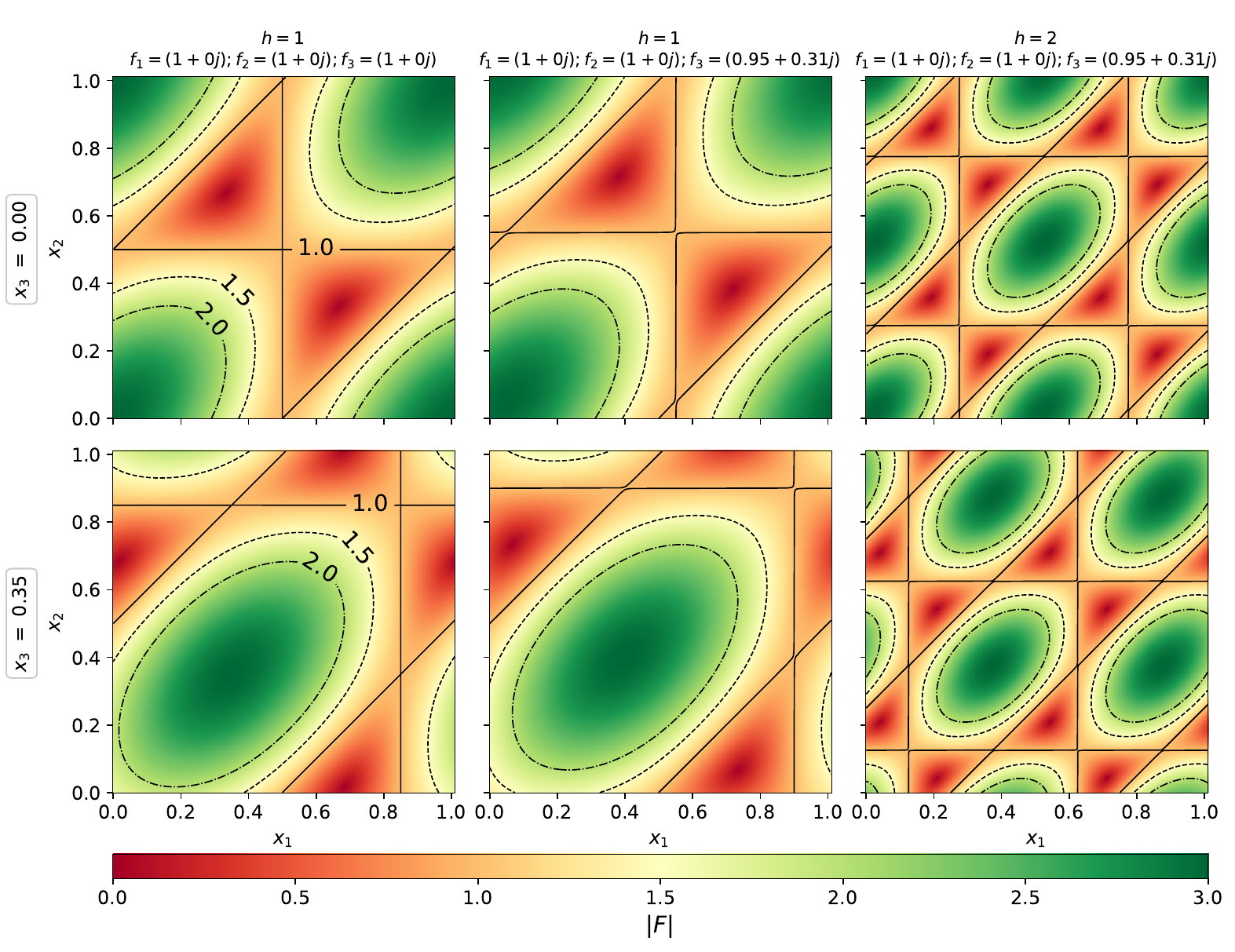}
    \caption{An example of typical isosurfaces in the parameter space concept, generated for an arbitrary 1-dim. acentric structure consisting of three atoms. The atomic coordinates $(x_1, x_2, x_3)$ are varied in the limits $[0, 1]$, \ie covering the full unitcell of the crystal structure. For simplicity, the coordinate of one atom, along the $x_3$ direction, is restricted at $x_3=0$ (top row) and $x_3=0.35$ (bottom row). The isosurfaces of magnitudes $|\isof|$ 1.00 (solid line), 1.50 (dashed line), and 2.00 (dash-dotted line) are highlighted for each combination of $f_i$ and $h$ (assumed values are given at top). The color map represents the calculated intensity for a combination of atomic coordinates within the range of 0 and 1. The color bar indicates the values of computed intensity.}
    \label{fgr:f2_clx}
\end{figure}

Of all possible types of isosurfaces, those based on $g(h)$, \ie $\isoG{h}{g(h)}$, are the most general ones and also geometrically simplest. They deserve, however, precise absolute scaling as all $\mathcal{M}$. Geometric properties are quoted for $\isoG{h}{g(h)}$ in Sec. 4.2 in Ref.~\cite{ZF09}. Inspecting the various distributions, \eg $g(h,\bm{t})$ over the full $\P{m}$, scaling symmetry is observed (in Ref.~\cite{FKZ05} un-precisely called self-symmetry), signifying that the landscapes of $\isog(1)$ isosurfaces are linearly reduced by the factor $\nicefrac{1}{h}$ in each dimension, and then repeated by symmetry until $\P{m}$ is complete, see Fig.~\ref{fgr:isosurfaces} and Fig.~3 in Ref.~\cite{FKZ05}. This scaling symmetry holds similarly for $\isog$ and $\isof$ (also for those based on intensity and amplitudes). The isosurfaces of centric and acentric 1-dim. structures are different, \eg in geometry and probability distribution of their amplitudes, see Fig.~\ref{fgr:f2_clx} and page 327, Figs.~1c,~2c in Ref.~\cite{KF09}.

\section{\label{sec:generalstrategies}Strategies for solving structures}
A comprehensive introduction to the PSC formalism further requires knowledge of finding the solution vector $\ta{s}$ that best matches the experimental observations. Here, a set of general strategies for solving crystal structures is presented. The discussion is organized into sections to introduce previous fundamental concepts for solution finding (see Sec. \ref{sec:solugeneral}), describe a grid-based direct-space approach, where $\ta{s}$ is determined by quantitatively comparing trial solutions with observed data (see Sec. \ref{sec:optionA}), outline the method that directly assesses trial vectors $\ta{a} \approx \ta{s}$ by progressively intersecting isosurfaces, thus narrowing down the solution space (see Sec. \ref{sec:optionB}), and to focus on reducing the solution region by applying a system of inequalities, providing qualitative agreement with the measurements (see Sec. \ref{sec:optionC}). These strategies offer flexible application of the PSC method for structure determination.
%\textit{how to obtain a model of the crystal structure containing atom coordinates (and perhaps other additional parameters) from Bragg scattering data?}

\subsection{\label{sec:solugeneral} General strategies}
PSC solves crystal structures from amplitudes or intensities of standard Bragg reflections without being affected by the phase problem or the necessity to de-convolute a Patterson function. However, much higher computing and storage demands are paid, as will be shown below in theory and discussed in practice, \ie numerically, in Sec.~\ref{sec:preference}. Such a new method cannot, of course, extinguish the more general problem of ``how to obtain a model of the crystal structure containing atom coordinates (and perhaps other additional parameters) from Bragg scattering data?'' -- in other words, ``how can the inversion of structure factor equations be done''. What it may can, is to provide a new picture of a structure model instead of the conventional distribution of scattering density in the unit cell (thus its denstiy peak coordinates as well as thermal smearing of maxima) or the (mathematically) corresponding set of numeric parameters obtained by a standard least-squares refinement of a nearly correct model~\cite{Coc48}. Hence, the general problem is looked at from another (mathematical) point-of-view and the result has other general features - both advantages and shortcomings, as shortly mentioned in the \textit{Abstract} of this review. It may describe the model by another selection of numerical parameters.

In the last years, essentially different methods were derived to find a (mainly geometric) model solution $\ta{a} \approx \ta{s}$, \eg a conventional least-squares refinement. We have introduced above (Sec.~\ref{sec:solutionfinding}) and reported in literature three Options A, B, and C for solving, \eg, a 1-dim. projection from a batch of $n \geq m$ independent amplitudes $\left|F_{\mathrm{obs}}(h00)\right|$, assumed scaled to $K=1$ (cf. Eq.~\eqref{eq:rho}). Option A is the simplest one, as it works without isosurfaces; hence it is the most tested one. Searching in direct space, \eg within the grid appraoch (Option A), FOM($\mathbf{t}$) is optimized by simply varying $\mathbf{t}$ within the available grid points, see Sec.~\ref{sec:optionA} and page 648 in Ref.~\cite{FKZ05} or page 386 in Ref.~\cite{FKZ08}. For this, the presented definitions of $G(h)$ (see Eq.~\eqref{eq:g_acentric-centric}) suffice. Option B uses a selected type of isosurface and tries to find its intersections. Option C is based on inequalities between, say, $\left|F(h_1)\right|$ and $\left|F(h_2)\right|$ and their respective isosurfaces. Thus, the permitted solution region is reduced until it appears small enough as desired, in favorable cases by each inequality. Options B and C need numerical descriptions of isosurfaces (or of their approximations), which is a task on its own. The following sections will describe these options in more detail.

\begin{figure}
    \centering
    \includegraphics[width=0.7\textwidth]{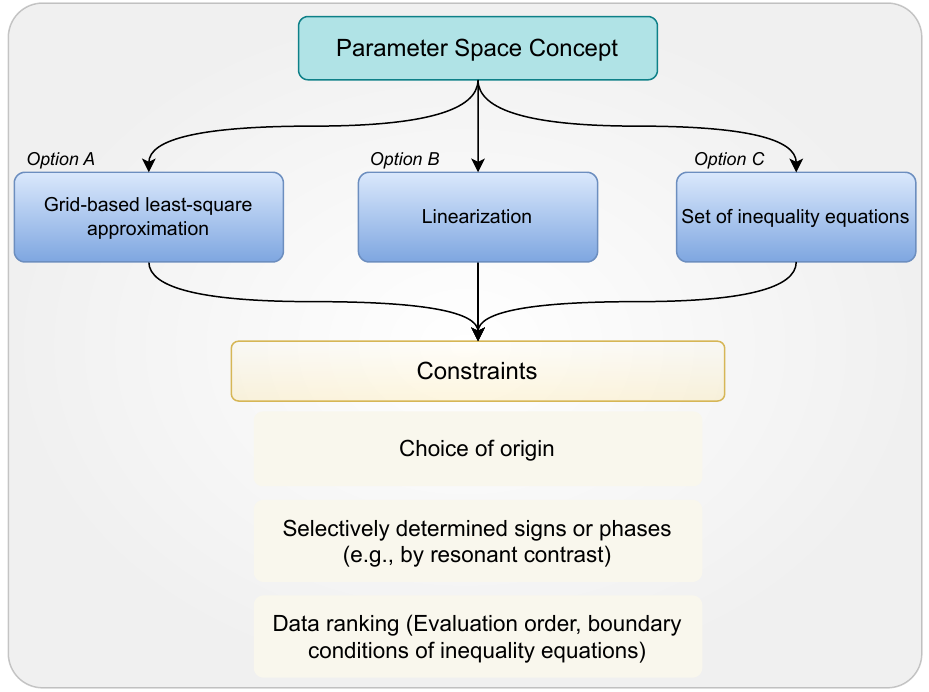}
    \caption{Basic explanation of structure solving options available in PSC. Each option may invoke further constraints to define the problem.}
    \label{fgr:pscmodels}
\end{figure}

% Searching in direct space, \eg within the grid appraoch (Option A), FOM($\mathbf{t}$) is optimized by simply varying $\mathbf{t}$ within the available grid points, see Sec.~\ref{sec:optionA} and page 648 in Ref.~\cite{FKZ05} or page 386 in Ref.~\cite{FKZ08}. For this, the presented definitions of $G(h)$ (see Eq.~\eqref{eq:g_acentric-centric}) suffice.
For these reasons, grid computing became the favored technique by Fischer \etal (for details see Sec.~\ref{sec:optionA}). They developed and inspected the other two approaches from a more academic view (see Sec.~\ref{sec:optionB} and Sec.~\ref{sec:optionC}). Using a standard PC and normal grid computing, good results were achieved at $m \le 11$ without paying much attention to optimizing the procedures~\cite{FKZ08}.

\subsection[Option A: Grid computing]{\label{sec:optionA}Option A: Grid computing, \ie direct space solution of $\ta{s}$ by quantitative comparison with observations}
The very first strategy developed within PSC for solving crystal structures was based on searching algorithms on structured grids, constructing minimization functions for each experimental observation, and solving them utilizing the least-squares technique. The following section of the grid approach is structured as (i) making general remarks to provide the methodologies (see Sec.~\ref{sec:gridgeneral}), (ii) focusing on theoretical approaches to standard grid scanning for various cases such as centric structures containing equi-point scatterers and selected acentric structures (see Sec.~\ref{sec:gridmethod}). Remaining challenges for the grid approach comprise further exploration of practical restrictions of complete grid searches, including grid errors and variations, coarse and fine grids, and special grid configurations. In particular, the concept of using varying gradients to refine the search process may provide further means to optimize specific grid characteristics.

\subsubsection{\label{sec:gridgeneral}General remarks}
Grid computing in its simplest form has three characteristics:
\begin{enumerate}[label=(\roman*)]
    \item It is a trial-and-error method providing mainly errors (\ie failures) and one or a few useful results $\ta{a}\approx\ta{s}$. In other words, it is a brute force technique.

    \item Although a single global minimum of FOM is aspired, this is not represented by just one $\ta{a}$, but by a small number of closely neighboring ones~\cite{Zschornak2020, ZW22}. As this is valid for all options, it is discussed generally in Sec.~\ref{sec:globalminimum} of the correspondence.
    
    \item Using not all of the grid's test vectors $\ta{t}$ in the asymmetric unit, but a subset selected at random, which can save much of the computing effort, leads also to an approximate solution $\ta{a}$, which can, however, not guarantee that $\ta{a}$ is the global minimum closest to $\ta{s}$. In case that high-dimensional solutions $m>20$ should come into closer reach, this random search technique would be even more advisable, due to the much smaller probability of ending up with a homometric problem.
    % Using not all \highlight[id=mn, comment={}]{[$^{TBR} mn->mv$ did you explain somehwere what this is? tt in general and asymmetric tt as a special case?] asymmetric $\ta{t}$}, but a subset selected at \highlight[id=mn, comment={}]{[$^{TBR} mn->mv$ This sounds like a very hand-wavy approach and undesireable from the outcome. It needs some smart restrictions. reference?] random, which can save much of the computing effort, leads also to an approximate solution $\ta{a}$, which can, however, not guarantee that $\ta{a}$ is the \textit{global minimum} closest to $\ta{s}$}. In case that high-dimensional solutions $m>20$ should come into closer reach, this random search technique would be even more advisable, due to the much smaller probability of ending up with a \highlight[id=mn, comment={}]{[$^{TBR} mn->mv$ what is a homometric problem?] homometric problem}.
\end{enumerate}

\subsubsection{\label{sec:gridmethod} Standard grid scanning} The key tasks of the grid approach are (1) to efficiently sample the asymmetric unit by predefined positional vectors, then (2) to calculate the targeted scattering property (geometric structure amplitude, structure amplitude, structure factor, intensity, etc.) for a selected set of reflections $hkl$, and then (3) to compare these values with the experimental observations. Steps (1) and (2) may be preprocessed, and the respective values stored in databases for a direct comparison within step (3) to speed up the PSC solution approximation. The sampling efficiency is crucial since the scaling of computations and storage is the limiting factor in treating high-dimensional degrees of freedom (see Sec.~\ref{sec:problem}).

For simplicity, we start with centric structures and geometric structure factors $G(h); h = 1,\ldots, 20$. Applying permutation symmetry, the asymmetric unit $\A{m}$ covers the $m$-dim. volume $[0 \le x_m \le x_{m-1} \allowbreak\le \ldots \le x_2 \le x_1 \le 0.5]$. For the grid construction, we take a structured grid with equal spacing in all $m$ dimensions $\Delta t=\Delta x_j=0.01, j=1, \ldots, m$. To save indexing storage, this grid structure allows for a defined sequence of positional vector incrementation, \eg with $x_1$ being the slowest (outer) loop, consecutively increasing the loop cycling with $x_j$, up to $x_m$ being the fastest (inner) loop. Each loop runs within the boundaries predefined by $\A{m}$. Using the database option, the necessary database storage is determined by the dimensionality $m$ (see Tab. \ref{tab:deltax} for specific computation as well as storage requirements), the number of reflections $h_{\mathrm{max}}$ (as a factor), as well as the stored digital precision of the scattering property. 

The digital precision of the data, as well as the grid spacing, directly reflect the possible resolution as well as the acquired precision and accuracy of the determined structures~\cite{ZW22}. In this study, model databases (steps (1) and (2)) have been computed for $m~=~2$ and $m~=~3~(h=1,\ldots, 20)$ and stored in files. The PSC solution space was then successfully identified (step (3)) by comparison to synthetic diffraction data~\cite{ZW22}.

\subsection[Option B: Direct assessment by step-wise intersecting isosurfaces]{\label{sec:optionB}Option B: Direct assessment of $\ta{a}\approx\ta{s}$ by step-wise intersecting isosurfaces}
\begin{figure}
    \centering
    \includegraphics[width=0.5\textwidth, clip, trim=0in 0in 0.4in 0.in]{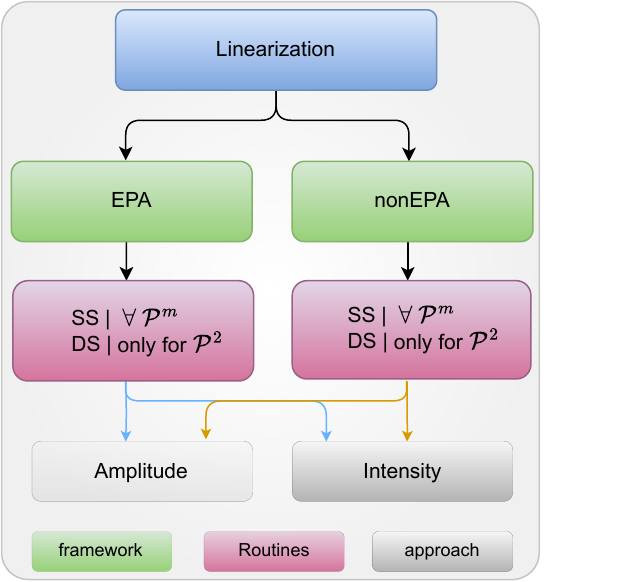}
    \caption{The linear approximation offers routines for single-segment (SS) and double-segment~(DS) treatment within both EPA and non-EPA frameworks. Note that the SS method is generally applicable to all PS of $m~\geq~2$, while DS so far is exclusively implemented for PS of $m~=~2$~\cite{code}. The elegance of treating realistic structures with non-equal atomic scattering factors $f(E, \mathbf{Q})$ (non-EPA), elevates the previous restriction of approximating all atoms solely based on the geometric structure}.
    \label{fgr:psclinear}
\end{figure}

Following the developments in grid scanning, the linear approximation (or equivalently linearization) has been implemented within PSC recently. The linearization is structured into frameworks, methods, and approaches, cf. Fig.~\ref{fgr:psclinear}. Depending on available experimental information, these options can be invoked to solve for both homometric and quasi-homometric structures in a single simulation. The routines handle the diffraction information in two approaches, amplitude (when phase is known) and intensity (when phase is unknown). The amplitude approach allows for solving with fewer ambiguities due to the known phase than the intensity approach, resulting in a less degenerate solution vector $\ta{s}$.

The frameworks offer options to treat the structure as geometric-approximated scatterers (the EPA framework), or as realistic scatterers (the non-EPA framework, where scatterers possess respective scattering strengths $f(E, \mathbf{Q})$; $E$ and $\mathbf{Q}$ are the energy of incident \xrays and the momentum transfer vector, see Eq.~(3) in Ref.~\cite{ZW22}). Additionally, the amplitudes approach offers the use of the sign of experimentally observed structure factors, thereby providing less degenerate and more precise solutions than the intensity approach, which is employed when such experimental information is unavailable. With these options, the linear approximation of experimental or calculated amplitudes/intensities is carried out in a stepwise process~\cite{code} as follows, cf. Fig.~\ref{fgr:fc}:
\begin{figure}
    \centering
    \includegraphics[width=0.98\textwidth]{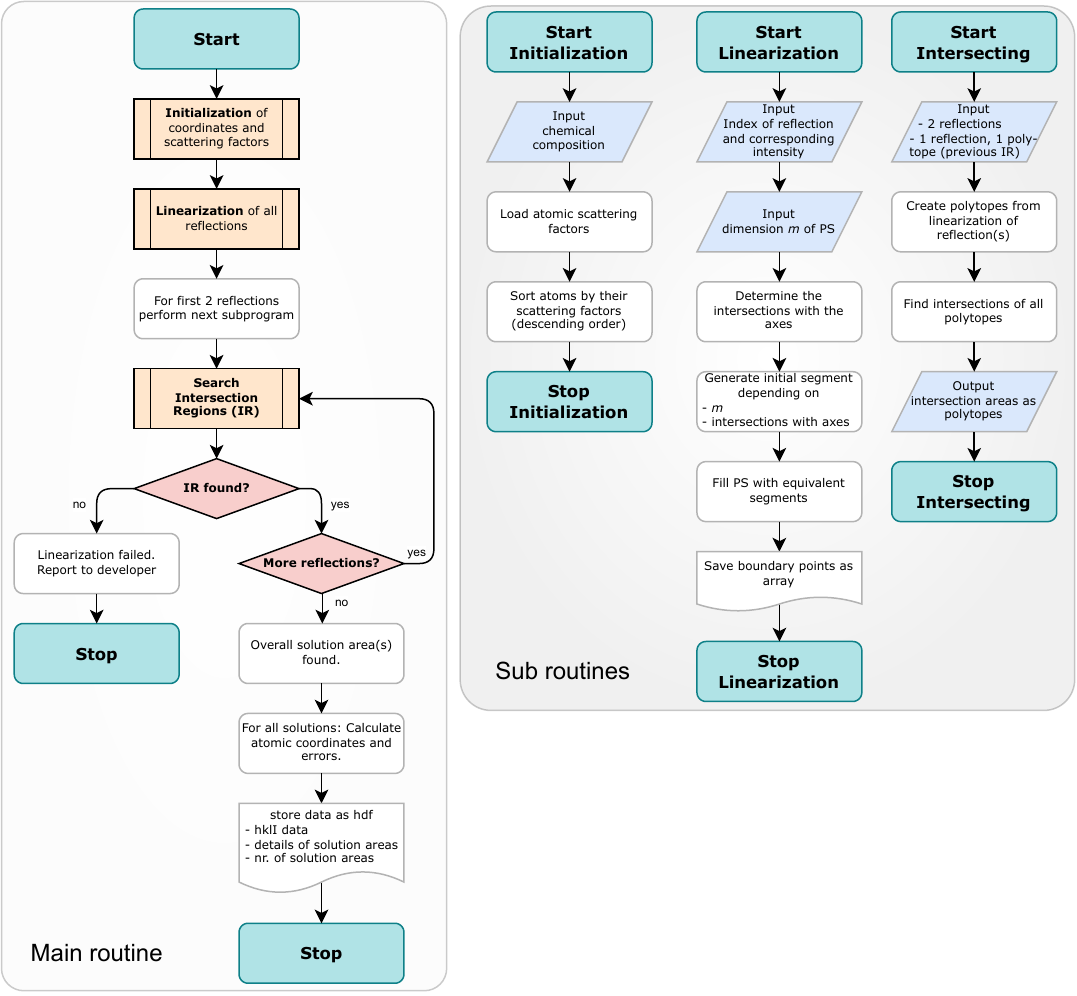}
    \caption{The flow of operation involved in linearization implemented within the PSC program \texttt{pyPSC}~\cite{code}. The operation comprises decision-making (red) with more reflections to include more data at the cost of computation time. The structure determination is terminated after processing the last available experimental (or theoretically calculated) amplitudes/intensities. The results, such as the determined structure solution/solutions, error on each solved atomic coordinate, performance parameters, etc., are saved in an HDF file. Recreated from Ref.~\cite{VNZ2024} with permission.} 
    \label{fgr:fc}
\end{figure}

\begin{enumerate}[label=(\roman*)]
    \item \textbf{Initialization:} In the first step, specific measures are implemented to ensure the flawless execution of approximating amplitudes/intensities. The measures involve ordering atomic structure factors $f_i$ from large to small, enforcing permutation symmetry from $m!$ if applicable~\cite{FKZ05}, and assigning them to orthogonal axes in $\P{m}$ with inverse index, persuasively predicting the general curvature of the isosurfaces~$\isoF{h}{|F(h)|}$. This ordering exerts maximum control over the behavior of $\isof$. While the lighter the atoms, the smaller the influence on the curvature of $\isof$ and leads to various topologies, see Sec.~2.4.1 in Ref.~\cite{VNZ2024} for more details. In addition, possible constraints are imposed, cf. Fig.~\ref{fgr:pscmodels}. 

    \item \textbf{Approximation:} Approximating the targeted solution space instead of locating the exact solution vector is realized here, mitigating the demerits of the grid-based approach, mainly the curse of dimensionality up to a certain degree. By means of linear boundaries, the solution space is restricted based on the amplitudes/intensity of each experimental observation by parallel lines in~$\P{2}$, planes in $\P{3}$, or hyper-planes in~$\P{m}$; $m~>~3$. These parallel boundaries basically present a set of linear inequalities defining the mathematical object of a polytope, effectively separating the allowed solution space from the complete PS for each experimental observation taken into account. The formulation of these boundaries is presented in Ref~\cite{VNZ2024}. Also, the linear approximation invokes the Mean-Value Theorem~\cite{MVT1} at the core, guaranteeing the existence of the approximation.
    
    \item \textbf{Determination of solution vector $\ta{s}$:}
    The inequalities, \ie the polytopes, contain as many variables as the dimension $m$ of PS, instead of $n^{3m}_\mathrm{step}$ (see Sec. 4.1) comparisons as for the dense grid-based approach (see Sec. \ref{sec:preference}), to be determined utilizing the set operations in order to perform the so-called intersection between various experimental observations. The intersection will result in a single or multiple (symmetrically equivalent as well as in-equivalent) polytopes. Only in ideal cases with error-free amplitudes for a full set of reflections starting from $h~=~1$, the solution vector $\ta{s}$ is uniquely represented, \ie, an unambiguous solution is obtained, whereas in general, as is inherent to PSC methodology, both the homometric and quasi-homometric solutions are determined simultaneously.
    
    \item \textbf{Data writing to HDF:}
    The solution vectors are so far represented as polytopes, but more informative parameters may be analyzed, \ie the volume of solution regions, the approximated structures, and the error on each solved coordinate. Those main results, along with performance parameters like the time consumption, are written into a Hierarchical Data Format (HDF), an ideal and versatile format for large and complex datasets~\cite{HDF}, for each experimental observation.
\end{enumerate}
The processes described above are integrated into Python routines to perform linearization within PSC~\cite{code}. Extensive tests in $\P{2}$ and $\P{3}$, carried out on Monte Carlo–generated artificial crystal structures and across a wide range of $f_i$ combinations within both EPA and non-EPA frameworks, are reported in Ref.~\cite{VNZ2024}. The $\ta{a}$'s are demonstrated as solution space within PS having a certain volume, as small as \num{e-6} of the unit cell volume, with extension in the order of \num{e-4} along each structural degree of freedom.

Though mitigating the memory problem of the grid scanning, where $\Delta t$ controls the accuracy demanding huge storage capacities, see Sec.~\ref{sec:gridmethod}, the linearization still possesses the following major demerits, which are yet to be tackled as futuristic goals.

\begin{enumerate}[label=(\roman*)]
    \item The routines in the operation flow, cf. Fig.~\ref{fgr:fc}, are not yet coded in parallel execution, but still executed in sequential order, demanding increased computing time.
    
    \item Selection of the least correlated and most efficient experimental observations, \ie those potentially exploring $\ta{s}$ with minimal volume in $\P{m}$, is not yet developed.
    
    \item The periodic nature of the isosurface~\cite{KF05} is both curse and fortune. It discovers deterministically (the fortune) all possible locations of $\ta{s}$ while requiring computational repetition (the curse) of the polytope creation in the complete PS, which is yet to be tackled in particular for the applications in PS of $m > 10$. Note that the number of repetitions in the complete PS is given by $2 \cdot h^m$. 
    
    \item For a further reduction and efficient storage of the final PSC results, it would be beneficial to implement structural symmetry analysis routines to separate homometric and quasi-homometric solutions.
    
    \item The error due to linear approximation of $\isog$ or $\isof$ may result in false-minima solutions. Also different segmentation methods of $\isog$ or $\isof$, single segment and double segment linearization \cite{VNZ2024}. The double segment is shown to perform better as evidenced by monitoring the volume of solution(s) enclosed by the polytope(s). Therefore, further reduction of false-minima solutions can be achieved by an increased number of segments and anchor points in the linearization process. Note that although the single segment approach is developed for general $\P{m}$ PS, the double segment approach is so far only for $\P{2}$.
\end{enumerate}

The linearization routines and their implementations within PSC offer a generally applicable alternative to established structure determination methods, like the widely used Fourier inversion. The ongoing efforts aim to enhance the efficiency of data handling and overcome current constraints, promising further advancements in the capabilities and accuracy of PSC.

\subsection{\label{sec:optionC}Option C: Reduction of solution regions by inequalities (qualitative agreement with observations)}
In the years before the DMs were developed, inequalities between selected reflection amplitudes/intensities were used to obtain sign relations for Fourier calculations. Although being the least developed option within the methodology, PSC inherently permits the use of any inequalities in principle, expressed \eg as  $\left|F(h_1)\right|<\left|F(h_2)\right|$ or $q(h_1, h_2):=\nicefrac{\left|F(h_1))\right|}{\left|F(h_2)\right|}<1$. Hence, a new type of isosurface $\mathcal{M}(q=1)$ divides $\A{m}$ into two volumes (regions): one cannot hold $\ta{s}$, and is thus forbidden. The permitted solution region can become smaller by the next inequality. Thus, by $n$ reflections in the data set, $n-1$ such independent inequalities can be applied, resulting in one or a few regions small enough to obtain estimated vectors $\ta{a}$ close to $\ta{s}$.

\subsection{\label{sec:constraints}Constraints}
The inherent symmetry of PS and model-based symmetry of isosurfaces contribute to reducing the complexity in the structure determination process. The free choice of fixing the origin in the complete PS and selecting the sign of the first encountered reflection, either based on experimental knowledge or theoretical calculations, is another constraint that directly limits the degrees of freedom to be solved by PSC. The detailed analysis of PS and isosurface symmetry can be found in Ref.~\cite{FKZ05}.

\subsubsection{\label{sec:originchoice} Inherent permutational symmetry, choice of origin}
The inherent symmetry of the parameter space in the case of EPA facilitates the reduction of the parameter space $\P{m}$ to an asymmetric unit $\A{m}$ through the application of permutation symmetry of atomic coordinates. For centrosymmetric structures, the reduction of $\P{m}$ to $\A{m}$ encompasses both positive and negative sign instances of the isosurfaces and fixes the vertices of $\A{m}$ at $[0,0,\dots,0],~[0.5,0,\allowbreak \dots,0],~\dots,~[0.5, 0.5, \allowbreak \dots,0.5]$, see the example case for $m=3$, Fig. 3 in Ref.~\cite{VNZ2024}. A limited permutation symmetry in the case of non-EPA with $n$ inequivalent scatterers reduces $\P{m}$ to $\A{m-n}$. In addition to the permutation of atomic coordinates, the origin of the crystal system is a free parameter for the first encountered reflection and can be fixed at one of the two centers of inversion (at $[0,0,\dots,0]$ and $[0.5,0,\dots,0]$) for centrosymmetric or anywhere in $\P{m}$ for non-centrosymmetric structures, which further reduces the possible volume of parameter space to search for the solution vector. The isosurface of zero amplitude, see yellow shading in Fig.~\ref{fgr:isosurfaces}, defines one boundary of the asymmetric unit. Once the choice of origin is applied for the first processed reflection, only the region below or above the zero isosurfaces needs to be analyzed, as per the known or pre-assumed sign of the respective isosurface.

\subsubsection[Selectively determined signs or phases]{\label{sec:resonant} Selectively determined signs or phases (\eg, by Resonant contrast)}
Although generally not directly available in conventional lab diffraction experiments, due to loss of the phases of structure factors $F$ by measuring $I(hkl)=|F(hkl|^2$, modern synchrotron experiments allow for direct phase determination by means of resonant contrast, which has been widely applied in Resonant Elastic \xray Scattering~\cite{panet01260} for accurate crystal structure refinements~\cite{Materlik1994, Joly2012, Zschornak2014, Richter2018, Weigel2024} and may even become the dominant signal in the case of purely resonant diffraction at forbidden reflections~\cite{Dmitrienko1983, richter2014mech, Ovchinnikova2020}. These phases may directly be used in PSC to further reduce the solution space, \eg, by a factor of two for each known sign in the case of centrosymmetry (see Ref.~\cite{VNZ2024}).

Resonant contrast further allows for identification of false pseudo-solutions, as a varying scattering power $f_i(E)$ of specific atoms $i$ will shift such false solutions in PS, whereas real solutions will maintain their solution vector $\ta{s}$ (see Sec.~\ref{sec:Resonant} 5.2). In addition, the deformation of individual isosurfaces induced by energy-dependent variations of $f_i(E)$ will significantly improve the resolution obtained by PSC. 

These features have been demonstrated by Zschornak \etal, applying diffraction datasets of two photon energies, below and above a specific adsorption edge, to resolve a split position in \larsmo, with sub-pm resolution~\cite{ZW22}.

\subsubsection[Data ranking]{\label{sec:EO} Data ranking (\eg, Evaluation order)}
Obtaining the $\ta{s}$ as early as possible in the PSC structure determination reduces computational load, in particular as the complexity of finding $\ta{s}$ increases with increasing $m$. This may be achieved by optimizing the order in which the reflection data is processed, which requires a clever sequence of reflections. Under this aspect, considering the Miller index $h$ of reflections with respective amplitudes or intensities, the most reasonable sequences that can be proposed are based on (i) low-indexed reflections that have the largest continuous regions in PS for given signs and thus less degeneracy at the cost of less precise solution regions, (ii) reflections with $g(h)$ or $|F(h)|$ equal or close to $m$ or $|\sum_i f_i|$, and (iii) the reflections with $g(h)$ or $|F(h)|$ equal or close to 0. On this basis, available reflection data may be ranked and named Evaluation Order (EO). Different EOs have already been discussed~\cite{FKZ05}, among them Natural Order, High Value Order, and Weighted High Contrast Order. The Natural Order selects the observations according to their regular order using Miller indeces (\ie  $1, 2, \ldots, q$, where $q$ is the number of available or assumed reflections). The High Value Order creates a new sequence of reflections, such that intensities $I(h)$ or respective amplitudes are sorted from highest to lowest value, \eg, as $g(h_1) > g(h_2) > \ldots > g(h_{q-1}) > g(h_q)$. The Weighted High Contrast Order normalizes the amplitudes or intensities by their corresponding Miller indeces $h$ and sorts the fraction highest to lowest, \eg, as $\nicefrac{g(h_1)}{h_1} > \nicefrac{g(h_2)}{h_2} > \ldots > \nicefrac{g(h_{q-1})}{h_{q-1}} > \nicefrac{g(h_q)}{h_q}$ and alternating the sequence with high and low values, \eg $\nicefrac{g(h_1)}{h_1}, \nicefrac{g(h_q)}{h_q}, \nicefrac{g(h_2)}{h_2}, \nicefrac{g(h_{q-1})}{h_{q-1}}, \ldots$. So far, only the natural evaluation order has been thoroughly tested~\cite{FKZ05, ZW22, VNZ2024}.

\section{\label{sec:problem}Problem: dimensions of (and in) Parameter Spaces}
In this section, the fundamental dimensionality-related challenges in PSC are examined. We address the computational preference for 1-dim. projections, exploring how these intuitive simplifications may obscure critical multidimensional relationships in Sec.~\ref{sec:preference}. Considering the notion of a global minimum in the structure determination context, emphasizing the conceptual and practical limitations of reducing the exact structure determination to a single optimal solution in Sec.~\ref{sec:globalminimum}. Also, the role of symmetry systems in PSC is discussed by analyzing how inherent symmetries within structures influence the navigation and interpretation of high-dimensional parameter spaces, see Sec.~\ref{sec:symmetry}.

\subsection{\label{sec:preference}Preference of 1-dim. projections}
While the concept of isosurfaces is applied to options B and C (Sec.~\ref{sec:optionB} -- \ref{sec:optionC}), the simplest version of Sec.~\ref{sec:optionA} deals with an $m$-dim. grid of grid-points with equal spacing $\Delta t$, \eg $\Delta x_j\approx0.01$, in praxis adjusted according to the inherent (rather high) resolution. The number $N$ of grid-points needed for a brute force technique, applied to a 3-dim. acentric structure, thus suffers from the curse of dimensions, well known in informatics:
\begin{subequations}
    \begin{align}
        N(3) &= \left(\frac{1}{\Delta t}\right)^{3m} = n_{\mathrm{step}}^{3m} \label{eq:N(3)}\\
\intertext{$N(3)$ is valid for each amplitude of $(hkl)$ considered. Correspondingly, for 2- and 1-dim. structures or structure projections, $N(2)$ and $N(1)$ are:}
        N(2) &= \left(\frac{1}{\Delta t}\right)^{2m} = n_{\mathrm{step}}^{2m} \label{eq:N(2)}\\
        N(1) &= \left(\frac{1}{\Delta t}\right)^{m} = n_{\mathrm{step}}^{m}. \label{eq:N(1)}
    \end{align}
\end{subequations}

Therefore, (i) the limitation to 1-dim. projections appear mandatory saving a factor of ${(1/\Delta t)}^{2m}$ and (ii) assuming equal atomic point scattering (EPA model, if applicable) results in a permutation symmetry of order $m!$, which offers further potential to reduce $\P{m}$ to its asymmetric part $\A{m}$ to refer always to the same one for different calculations. Hence, of all $m!$ the one with corners $[0, 0, \ldots, 0, 0], [1, 0, \ldots, 0], [1, 1, \ldots, 1, 0], [1, \ldots, 1]$ is designated as $\A{m}$. It contains the $m$ independent atoms having coordinates  $x_1 \geq x_2 \geq \ldots \geq x_m$ (cf. p.~645 -- 646 (Figs.~1 and 2) in Ref.~\cite{FKZ05} or p.~447 in Ref.~\cite{ZF09}). For centric projections replace $n$ by $\nicefrac{n}{2}$ in Eq.~\eqref{eq:N(1)} and $1$ by $\nicefrac{1}{2}$ for the corners).
Applying the permutation symmetry to $N(1)$ fosters
\begin{equation}
    N_A\left(1\right) \approx \left(\frac{1}{\Delta t}\right)^m = \frac{n_\mathrm{step}^m}{m!}, 
    \label{eq:NA}
\end{equation}
thus approaching bearable numbers of structure factor calculations for each reflection. For some $m$ and assumed $\Delta x_j\approx0.01$, thus $n_{\mathrm{step}} = 100$ (for centric structures $n_{\mathrm{step}}=50$), see Tab.~\ref{tab:deltax}.

Two remarks must be made:
\begin{enumerate}
    \item While the $m$-dim. volumes of $\P{m}$ and $\A{m}$ are related by factorial ($!$), their numbers of grid points $N_P$ and $N_A$ are not, because $N_A$ also holds some more $(m-1)$-, $(m-2)$-, $\ldots$, $1$-, $0$-dimensional points at surfaces, edges and end-points. Thus, $m!$ in Eq.~\eqref{eq:NA} is only an approximation, although sufficient for estimating computer demand as shown in Tab.~\ref{tab:deltax}.
    
    % \item If not \highlight[id=mn]{all $m$ [What does it mean? Why are there several m?]} are equal or nearly so (as is assumed, say, for organic molecules also by DM), but merely $m\replaced[id=mn]{\prime}{'} \le m$, the computing demand is reduced only by a factor $\approx m\prime!$, perhaps also by $(m-m\prime)!$.
\end{enumerate}
\vspace{0.5cm}

\begin{table}
    \caption{Number of structure factor calculations for the grid-based PSC approach. Total required computations are given for centric as well as acentric structures with grid spacings $\Delta x_j=0.01$ and $\Delta x_j=0.02$ for several dimensions $m$$^\dag$.}
    \resizebox{\textwidth}{!}{
        \renewcommand{\arraystretch}{1.0}
        \begin{tabular}{c | c | c | c | c}
            \hline
            \hline
            
            & \multicolumn{2}{c|}{Centric} & \multicolumn{2}{c}{Acentric}\\ [1ex]
            \hline
            dimension ($m$) & $\Delta x_j=0.01$ & $\Delta x_j=0.02$ & $\Delta x_j=0.01$ & $\Delta x_j=0.02$\\ [1ex]
            \hline
            2  &    1250	    &   313	        &   5000	    &   1250       \\ %[1ex]
            3  &    20833	    &   2604	    &   166667	    &   20833      \\ %[1ex]
            4  &    260417	    &   16276	    &   4166667	    &   260417     \\ %[1ex]
            5  &    2604167	    &   81380	    &   83333333	&   2604167    \\ %[1ex]
            6  &    21701389	&   339084	    &   1388888889	&   21701389   \\ %[1ex]
            8  &    968812004	&   3784422	    &   2.4802E+11	&   968812004  \\ %[1ex]
            10 &    2.6911E+10	&   26280708	&   2.7557E+13	&   2.6911E+10 \\ %[1ex]
            12 &    5.0969E+11	&   124435168	&   2.0877E+15	&   5.0969E+11 \\ %[1ex]
            14 &    7.0012E+12	&   427318573	&   1.1471E+17	&   7.0012E+12 \\ %[1ex]
            16 &    7.2929E+13	&   1112808784	&   4.7795E+18	&   7.2929E+13 \\ %[1ex]
            18 &    5.9583E+14	&   2272893757	&   1.5619E+20	&   5.9583E+14 \\ %[1ex]
            20 &    3.9199E+15	&   3738312101	&   4.1103E+21	&   3.9199E+15 \\ %[1ex]
            \hline
            \hline
        \end{tabular}
        }
        {\footnotesize { $\dag$ A. Kirfel and K. Fischer realized that up to the order $N_A(1)\approx10^{14}$ can be handled by a standard PC.}}
        \label{tab:deltax}
\end{table}

\subsection{\label{sec:globalminimum}Global minimum of correspondence}
Assuming that the 1-dim. true structure $\mathbf{t}_s$ consisting of $m$ atoms does not lie exactly on a point $\mathbf{t}_t$  of the grid currently in use. It follows that a hypothetical geometrical figure of merit $\mathrm{FOMX}$ equals
\begin{equation}
    FOMX = \left|\ta{t} - \ta{s}\right| = \Delta t,
    \label{eq:FOMX}
\end{equation}
where the $2^m$ nearest $\ta{t}$ have a $\Delta t \le \Delta$ of this grid. Further, any observational figure of merit (FOM), \eg the agreement factor FOM for a sufficient number of models $\ta{t}$ FOM$(\ta{t})$ equals
\begin{equation}
    FOM(\ta{t}) = \frac{\sum_{hkl} \Bigl| K \cdot \left|F_\mathrm{obs}(hkl)\right| - \left|F_\mathrm{cal, \mathbf{t}}(hkl)\right| \Bigr|}{\sum_{hkl} \left|K\cdot F_\mathrm{obs}(hkl)\right|}
    \label{eq:FOM}
\end{equation}
($K$ = absolute scaling factor). 

Both can be ranked in descending order allowing for the identification of the models with the best agreements. While $\mathrm{FOMX}$ is useful only for test calculations on a given structure, $\mathrm{FOM}(\ta{t})$ can universally be applied to unknown structures. It may also be replaced by any other reliability index for comparing observed and calculated reflection amplitudes/intensities. The ideal $\mathrm{FOM}(\mathbf{t})=0$ can generally be only approximated. The distribution of both $FOM$'s is, however, not identical, but slightly different. This fosters the strategy to find, rank, and inspect a (small) number of possible solution vectors with the smallest $\mathrm{FOM}(\ta{s})$  rather than to consider only one $\ta{s}$ at the bottom of the ranking list.

\subsection{\label{sec:symmetry}Symmetry systems in PSC}
For our preferred 1-dim. projections, we have two symmetries: centric and acentric, thus also two types of structure factors for the general Wyckoff equipoint positions. For the two special positions (centric), special PSC handling is required. For two-dimensional structures (or projections), the 17 plane groups have different formulae for structure factors (including also centered unit cells); thus, they cannot be handled under the same two categories as the one-dimensional ones. In addition to various special positions (with specific eigen-symmetries), cell centering must also be taken into account. The situation is even more complicated in 3-dim. structures. 

Fischer \etal did not touch this challenge so far, besides the special positions in 1-dim. centric projections, see p.~645 (left column) in Ref.~\cite{FKZ05}. There, the case (AB) deserves handling by trial-and-error. Consequently, at present, 1-dim. structure projections still appear preferable, even if higher-dimensional structures should perhaps come into reach thanks to better computing facilities in the future.  

\section{\label{sec:scalefactor}Absolute scale factor and data reduction}
Properly processing the data is an essential practice for preventing incorrect structural results. So far, the parameter space concept has been tacitly presented on the assumption that the structure amplitudes or geometrical amplitudes, which must be derived from the observed relative reflection intensities lie on an absolute scale. In practice, this is generally not the case. The multiplication of the amplitudes with a correct scale factor is, however, an indispensable prerequisite, because the positions (and shapes) of the isosurfaces are directly affected by scaling. Though the scale factor can principally be determined experimentally (with a lot of effort and at best with a ca. 1--2\,\% relative uncertainty, \eg, by means of dedicated synchrotron XRD measurements), this is hardly practical. The alternative option, Wilson's statistics~\cite{Wilsona00174}, is not recommendable, particularly if only a rather small number of data is available (besides eventually picking up large and partly systematic errors). Different strategies have been devised specifically for PSC in past development phases. The scale factor problem can be circumvented by defining two quantities that preserve the intensity ranking of all reflections as well as the individual intensity fractions of the total measured intensity. These are addressed in terms of absolute scaling strategies according to amplitude or intensity ratios (Sec. \ref{sec:gorIratio}) and quasi-normalized amplitudes (Sec. \ref{sec:quasig}). 

\subsection[Amplitude or intensity ratios]{\label{sec:gorIratio} Amplitude or intensity ratios $q(\mathbf{h,k})$}
One measure for eliminating scaling that has been used successfully for a long time (and recently also in partial structure analyses) is using ratios of observations, \ie of intensities or amplitudes, recorded in one and the same experiment with its own scale factor. These ratios, each expressing the absolute contrast of scattering between any two reciprocal lattice points, can be treated as observations on the absolute scale, which, for a promising $\ta{t}$, should comply with the corresponding model amplitude relations.

Let the ratios be
\begin{equation}
    q_{\mathrm{obs}}(h_1,h_2) \equiv \frac{\left|Y(h_1)\right|}{\left|Y(h_2)\right|},\label{eq:q(h1,h2)}
\end{equation}
where $h_1$ and $h_2$ are elements of the batch of, say, $h00$ reflections and $Y(h)$ can be $K\cdot \left|F(\mathbf{h})\right|$ or $K^2\cdot\left|F(\mathbf{h})\right|^2$ ($\mathbf{h}$ in short for $hkl$) or any quantity derived from them (see Eq.~(2) in Ref.~\cite{KFZ06}). With the scale factor $K$ eliminated, $q(h_1,h_2)$ can be used in different options (for $n$ amplitudes $|F|$ or $g\prime$)
\begin{enumerate}
    \item using the largest one as the denominator for all $q$ or
    \item ranking the observations and using the larger of the two adjacent observations as the denominator for any $q$, 
    \item obtaining an anti-symmetric square matrix of $(n^2-n)$ $q$'s with $q\left(h_1,h_2\right)=1/q(h_2,h_1)$ and $n$ diagonal elements of 1. This follows from calculating all possible $q$'s in the batch of observations ranked.
\end{enumerate}

This results in $n-1$ independent observed ratios from $n$ reflections, which can be compared with the same group of corresponding model ratios $q_{\mathrm{cal}}(h_1,h_2)$ for each test vector $\ta{t}$. Clearly, $n-1$ must exceed or be equal to $m$. For $m = 4$, \eg $n=5$ suffices (apart from considering cases $q=0$). If $q_{\mathrm{obs}}(h_1,h_2) > 1$, one can substitute the ratio by its inverse to avoid too large figures. This way, $0 \le q\left(h_1,h_2\right) \le 1$ for all $q$. Quantification of agreement with the corresponding model data can be determined using Eq.~\eqref{eq:g_acentric-centric} where $q$ substitutes $g$ or $|F|$. If a $\mathrm{FOM}(q)$ is desired for any $\ta{t}$ (\eg in analogy to Eq.~\eqref{eq:FOM}, summation is over $n-1\geq m$. An intelligent selection of $q$, \eg $q\big[g(10)/g(10\pm1)\big]$ with high indexed $h_1$ and small $\Delta h$ to $h_2$, permits high nonius resolution (\ie angular resolution) at a stage already advanced for solution (see Fig.~4 in Ref.~\cite{KFZ06}). Each of the observed ratios $q-\left(h_1,h_2\right) \neq 1$, \eg of unscaled $g^\prime(\mathbf{h})$'s with $g^\prime(\mathbf{h})= K\cdot g_{\mathrm{exp}}(\mathbf{h})$, defines of course also an inequality of the scaled ones, \eg $g\left(h_1\right)\le g\left(h_2\right)$, see Sec.~\ref{sec:optionB}.

\subsection[Quasi-normalized Amplitudes]{\label{sec:quasig}Quasi-normalized amplitudes $e(\mathbf{h},n)$}
A second option avoiding the scale factor problem relates each observed intensity, $K^2\cdot{\left|F_{\mathrm{obs}}(\mathbf{h})\right|}^2$ or $g^{\prime 2}_{\mathrm{obs}}(\mathbf{h})$, to the intensity averaged over the $n$ reflections of the reflection batch used. The result is a list of $n$ so-called quasi-normalized structure amplitudes $e$ which are, by definition, on the absolute scale. This approach has been outlined in Ref.~\cite{KF09}, where the definition of quasi-normalization, the characteristics, and the consequences are treated in detail. Here, it may suffice to report the principal results. Quasi-normalized intensities are defined as

\begin{subequations}
    \begin{align}
        e^2(\mathbf{h}, n) &= n \cdot \frac{ g^{\prime2}_{\mathrm{obs}}(\mathbf{h})}{\sum_{k=1}^n g^{\prime2}_{\mathrm{obs}}(\mathbf{k}) } \label{eq:eg}\\
\intertext{or}
        e^2(\mathbf{h}, n) &= n \cdot \frac{ F^2_{\mathrm{obs}}(\mathbf{h})}{\sum_{k=1}^n F^2_{\mathrm{obs}}(\mathbf{k}) } \label{eq:eF}
    \end{align}
    \label{eq:e}
\end{subequations}
are obviously independent of the actual scale factor. Contrary to the use of $q$, all experimental values are explicitly retained upon calculating $e$ (Please note that in Eqs.~(2) of Ref.~\cite{KF09}, three types of $e(\mathbf{h},n)$ were defined, where only $e(\mathbf{h},n)$ (\ie Eq.~2c) is identical with the above $e$. In Ref.~\cite{KF09} $g^\prime(k)$ must be replaced by $g^{\prime2}(k)$ in Eq.~2c). 

As with $q$, comparison of experimental $e_\text{exp}$ data with the corresponding model quantities affords the calculation of $n$ model structure amplitudes followed by normalization to the corresponding $e_\text{cal}(\mathbf{h})$ for every $\mathbf{t}$.

Generally, the isosurfaces $\mathscr{E}(e,\mathbf{h},n)$ of $e(\mathbf{h},n)$ are not only geometrically different from the $\mathcal{G}(\mathbf{h},g)$ but also more complicated. This is because the denominator in Eq.~\eqref{eq:e} contributes its own $n$-dependent structure to the parameter space, cf. Fig.~3 in Ref.~\cite{KF09}. Thus, the characteristic properties of $\mathcal{G}(\mathbf{h},g)$ do not fully apply to $\mathscr{E}(e,\mathbf{h},n)$, cf. Figs.~4 and 7 of 1-dim. parameter spaces for centric $m=2$ and acentric $m=2+1$, respectively. In particular, ``downscaling'' as for $\mathcal{G}(\mathbf{h},g)$ compared to $\mathcal{G}(1,g)$ is no more possible, the $\mathscr{E}$-isosurfaces depend on $n$ as does, of course, the gradient field (Fig.~12 in Ref.~\cite{KF09}). Also, the topologies of the $\mathscr{E}$ and $\mathcal{G}$ landscapes differ in that the former may exhibit local maxima and minima in addition to the sites well known for $\mathcal{G}(\mathbf{h},g)$. 

The $e(\mathbf{h},n)$ and $g(\mathbf{h})$ become, however, more and more proportional with increasing $n$ because the modulation of the denominator (Eq.~\eqref{eq:e}) ceases accordingly (cf. Wilson statistics~\cite{Wilsona00174} and Ref.~\cite{KF09}; for regions close to lower-dimensioned boundaries of $\A{m}$ see Fig. 3c there). Hence, it follows that the geometries of $\mathscr{E}(e,\mathbf{h},n)$ isosurfaces (not the quantities $e$) tend towards those of the $\mathcal{G}(\mathbf{h},g)$ if only the data batch is sufficiently extended.

If a first 1-dim. projection has been solved and refined to satisfaction, the scaling of all reflection amplitudes of the same experimental batch is known. It may thus be transferred to other data sub-sets (including, of course, higher-dimensional ones) so that $g(\mathbf{h})$ could be used instead of $e(\mathbf{h},n)$. This offers little benefit for $\ta{t}$ based methods (Option A, cf. Sec.~\ref{sec:optionA}), but isosurface-based solution techniques (Option A and B, cf. Secs.~\ref{sec:optionA}-\ref{sec:optionB}) may gain due to the geometrically simpler system of $\mathcal{G}(h,g)$ as compared to that of $\mathscr{E}(e,\mathbf{h},n)$.

The use of $e(\mathbf{h})$ as well as their relation to the normalized structure amplitudes $|E(\mathbf{h})|$ of the DM and also the $e$-dependence of the batch size and composition were discussed in detail in Sec. A bridge to Direct Methods in Ref.~\cite{KF09}. Referring to Eqs.~5(b,c) of Ref.~\cite{KF09}, for a structure of $m$ randomly distributed equal point scatterers of scattering power $f=1$: 
\begin{subequations}
    \begin{align}
        g^2(\mathbf{h}) &= 2m \cdot |E(\mathbf{h})|^2 \quad \text{(centric)} \label{eq:centricg} \intertext{or}
        g^2(\mathbf{h}) &= m \cdot |E(\mathbf{h})|^2 \quad \text{(acentric)} \label{eq:acentricg} \intertext{and hence}
        e^2(\mathbf{h}, n) &= n \cdot \frac{|E(\mathbf{h})|^2}{\sum_{k=1}^n |E(\mathbf{k})|^2}. \label{eq:e2}
    \end{align}
    \label{eq:eforEPA}
\end{subequations}

Accordingly, $e(\mathbf{h},n)$ preserves the ranking of $|E(h)|$, \eg in descending order of magnitude, and since a large $e(\mathbf{h},n)$ is only possible in certain regions of $\A{m}$, it is not surprising to find that the combinations of the allowed regions associated to the largest $e$'s provide efficient reductions in the parameter space. The analogy to the fact is: phase relations are better defined the larger the products of the involved $|E|$'s.

The largest experimental $e$ values follow from, \eg, Figs.,~1 and 2 of Ref.~\cite{KF09}:
\begin{equation}
    e_{\mathrm{max}}(n)~\approx
    \begin{cases}
        2.7~\text{(centric) or} \\
        2.3~\text{(acentric) or}\\
        \sqrt{n}~\text{for}~n<7.
    \end{cases}
    \label{def_evalues}
\end{equation}
These numbers are almost independent of (not too small) $n$ and can be used for coarse scaling. 

\section{\label{sec:applications}Applications and feasibility studies}
This section presents practical applications and feasibility studies that demonstrate the relevance and implementation of the discussed theories of the PSC. Firstly, the resolution limits inherent in the approach are assessed via exploring how the precision of diffraction data influences structural prediction and interpretation, see Sec.~\ref{sec:resolutionstudy}. Following that, the general applicability for neutron scattering techniques is presented to explore further flexibility or constrain the proposed models, see Sec.~\ref{sec:nscattering}. Finally, the recent addition to PSC, linear approximation, is utilized to solve structures, simplifying complex parameter spaces as well as evaluating both their advantages and limitations in real-world scenarios, see Sec.~\ref{sec:linearization}.

\subsection{\label{sec:resolutionstudy}Resolution study}
Similar to the DMs, but in contrast to the optical picture of a crystal structure by Fourier inversion (\eg the \xray microscope), the spatial resolution in PSC is not limited by $\lambda/\sin{\theta_{\mathrm{max}}}$ of a complete data set (see weak points (ii) and (iii) in Sec.~\ref{sec:introgeneral}). A feasibility study of a 1-dim. centric projection showed how a better resolution with in PS can be obtained via PSC from a few Bragg reflection amplitudes (as a rule of thumb: number of reflections $\geq$ number of inequivalent atoms in the unit cell~\cite{VNZ2024}) applied to a limited core question~\cite{KF05, ZW22}, where a split position of two atoms was treated, resulting in a resolution down to the sub-picometer. The possible resolution depends on the quality of reflection measurements and independence of selected data (\ie, being free of correlation effects) range~\cite{ZW22}. 

\begin{figure}
    \includegraphics[width=\textwidth]{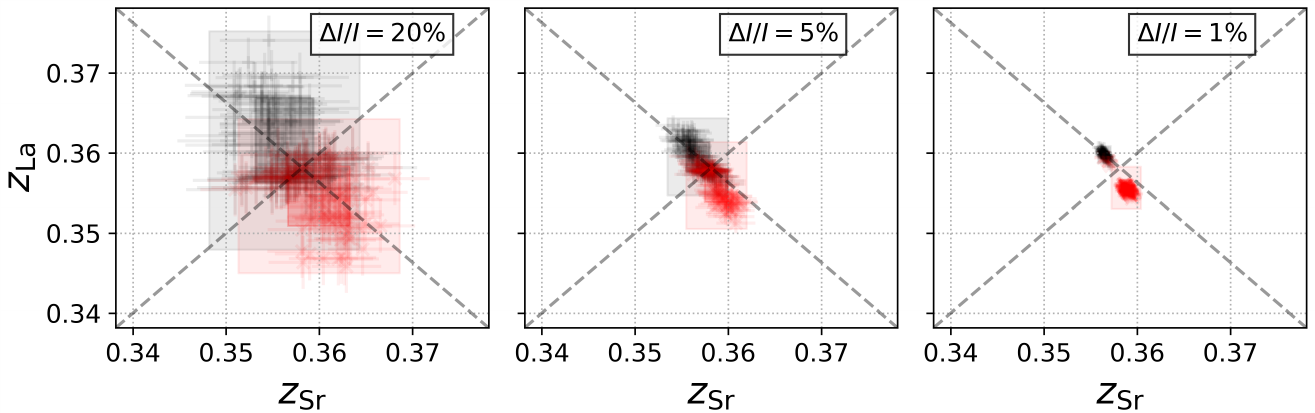}
    \caption{Monte-Carlo study to resolve the La-Sr split position in \larsmo with PSC. The reflection intensities are introduced with a Gaussian distribution-based random error for the reflections $h = 2, 4, \ldots, 20$. For each intensity error, almost \num{100} number of test samples are generated. The resolved solution (the black marks) and pseudo-solutions (red marks) are shown with confidence regions $1\sigma \approx \SI{68}{\percent}$ (darker shaded box) and $2.6\sigma \approx \SI{99}{\percent}$ (lighter shaded box). Results are copied from Ref.~\cite{ZW22} with permission.}
    \label{fig:4}
\end{figure}

To test the effect of XRD data quality on the structure determination, the presumed split position of La and Sr atoms in the potential high-temperature superconductor \larsmo has been investigated with respect to the error on reflection intensity by means of Monte-Carlo calculations~\cite{Kirfel2005, Zschornak2020, ZW21}. The La/Sr-split of $\Delta z = 0.0034$ ($\approx \SI{4.2}{\pico\metre}$) has been investigated using different error distributions of theoretically determined intensities $\Delta I/I$ of \SIlist{20; 5; 1}{\percent}, cf. Fig.~\ref{fig:4}. The error ranges result in different behavior. For the low-quality data with $\Delta I/I = \SI{20}{\percent}$, only one solution vector is allocable within the $2.6\sigma$ and $1 \sigma$ confidence regions in PS~\cite{ZW21}. In contrast, the high-quality intensities with $\Delta I/I = \SI{1}{\percent}$ resolve two distinct solution vectors, the true structure and a pseudo-solution (cf. Fig.~\ref{fig:4}), which both confirm the La-Sr split position. The split parameter along the $z$~axis has been well validated to be $0.0034 \pm 0.0003$ within the PSC procedure.

\subsection{\label{sec:Resonant}Resonant contrast}
\begin{figure}
    \begin{minipage}{7cm}
        \begin{overpic}[width=\textwidth]{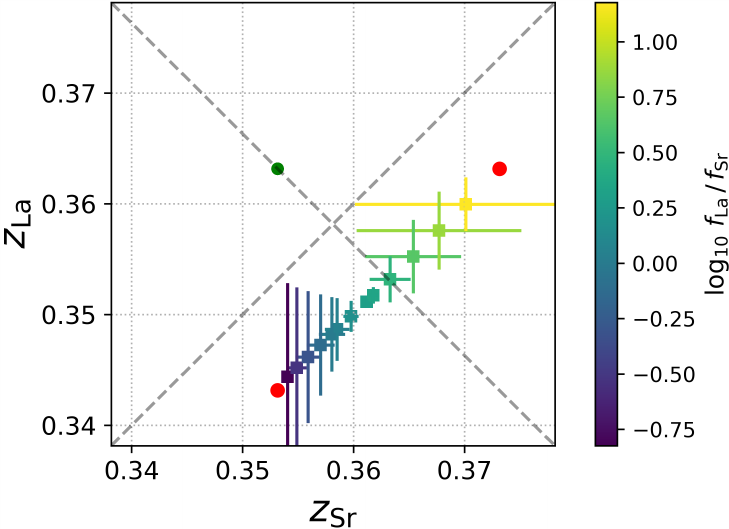}
        \put(0,70){(a)}
        \end{overpic} 
    \end{minipage}\hspace*{.5cm}%
    \begin{minipage}{7cm}
        \begin{overpic}[width=\textwidth]{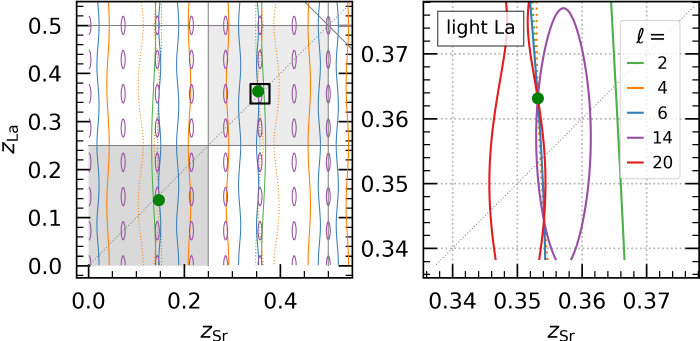}
        \put(-6,45){(b)}
        \end{overpic}
        \vspace*{0cm}%
        \begin{overpic}[width=\textwidth]{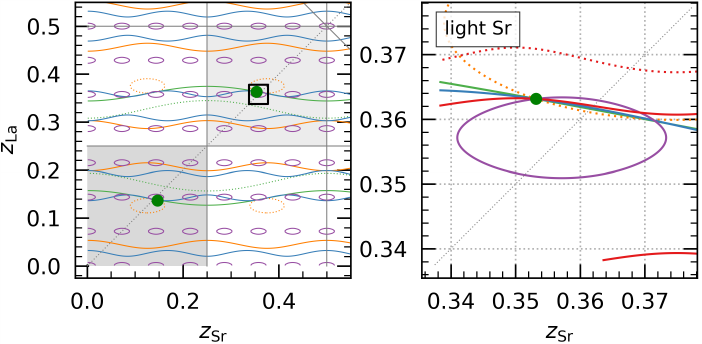}
        \put(-6,10){(c)}
        \end{overpic}
    \end{minipage}
    \caption{(a) Investigation on the true and the pseudo-symmetric solution as a function of scattering strength ratio by fitting intersecting isosurfaces $\isof$ calculated for reflections $l = 2, 4, \ldots, 20$. (a) Keeping $f_{\mathrm{La}} \cdot f_{\mathrm{Sr}}$ fixed, the ratio $f_{\mathrm{La}} /f_{\mathrm{Sr}}$ is varied by a scaling factor of 10. The error bars of $2.6\sigma$ denote the confidence regions, magnified by a factor of 10. The real solution $\ta{s}=(z_{\mathrm{La}}, z_{\mathrm{Sr}})$ is independent of $f_{\mathrm{La}} \cdot f_{\mathrm{Sr}}$ variation (green dot), whereas the pseudo-solutions result in a linear series between the limits at a distance $2 \cdot \Delta z$ (red dots). The change in scattering power changes the respective isosurface features, such as the elongation along the smaller scattering power axis. The ``light La'' (with artificially decreased atomic scattering factor $f_{\mathrm{La}}$) and ``light Sr'' effects are given in (b) and (c), respectively. Results are copied from Ref.~\cite{ZW22} with permission.}
    \label{fig:3}
\end{figure}

To evaluate the effect of resonant scattering contrast on positional resolution of the true solution and the pseudo-solution, Zschornak \etal varied the ratio of scattering strength $f_{\mathrm{La}}/f_{\mathrm{Sr}}$ while keeping the product $f_{\mathrm{La}} \cdot f_{\mathrm{Sr}}$ fixed, using simulations. Within the explored range, the coordinates of the true solution vector $\ta{s}$ remain unchanged, as they represent the fixed atomistic positions in the crystal structure. In contrast, the pseudo-solution shifts significantly in relative coordinates (Fig.~\ref{fig:3}(a)), with the displacement scaling with the scattering power variation. 

This behavior is typical for any pseudo-solution reflecting a structural pseudo-symmetry where a false solution lies close to the true one. The study directly modified the element-specific weights of the partial structure distribution, therefore the isosurfaces of a fixed amplitude (Fig.~\ref{fig:3}(b,c)) could only be preserved for the pseudo-solutions through a positional shift in parameter space, while the true solution anchors each isosurface. This observation suggests a general strategy to separate PSC solution volumes and identify false solutions by employing datasets at two photon energies and thus with different atomic scattering strength ratios, akin to resonant contrast in Resonant X-ray Diffraction.

The merit of reaching exceptionally high theoretical precision in PSC, particularly when combined with resonant contrast enhancement (see Ref.~\cite{ZW22}), assists in resolving challenging structural subtleties or pseudo-symmetries. While incorporating resonant enhancement into conventional Fourier inversion is not straightforward, the PSC framework offers the following distinct advantages: Correlations between parameters become directly perceptible through isosurface intersections, the remaining solution space quantifies positional uncertainties, and resonant contrast can be seamlessly integrated as additional diffraction datasets. The two-dimensional case presented by Zschornak \etal~\cite{ZW22} for \larsmo demonstrates the general applicability in the PSC framework as well as the obtained resolution enhancement down to the sub-pm range, and comparable resolutions can be anticipated for broader structural problems.

\subsection{\label{sec:nscattering}Neutron scattering}
In principle, neutron Bragg scattering amplitudes fit to PSC better than corresponding \xray data because the former are based on point-like scatterers, as does the PSC model. Anisotropy is thus mainly due to thermal smearing or disorder. It is well known that these two effects (i) exhibit non-negligible correlation coefficients with $\theta$-dependent \xray atomic scattering functions and (ii) can be separated by experiments at different temperatures. Reasons for amplitude and phase differences ($\Delta b$ and $\Delta\varphi$) in neutron scattering depend on the individual isotopes of a chemical species. As an example, for H and D isotopes $\Delta\varphi=\pi$. Furthermore, the incoherent scattering adds to the problem. Few nuclei exhibit resonant scattering, however, without absorption edges, as with \xrays.

\begin{figure}
    \centering
    \includegraphics[width=1\textwidth]{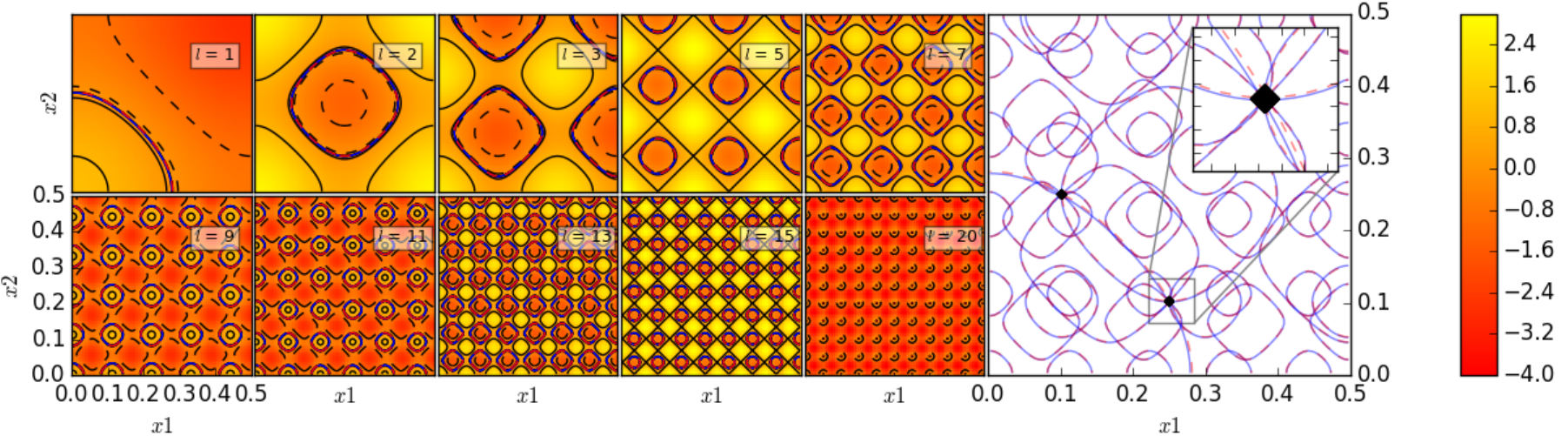}
    \includegraphics[width=1\textwidth]{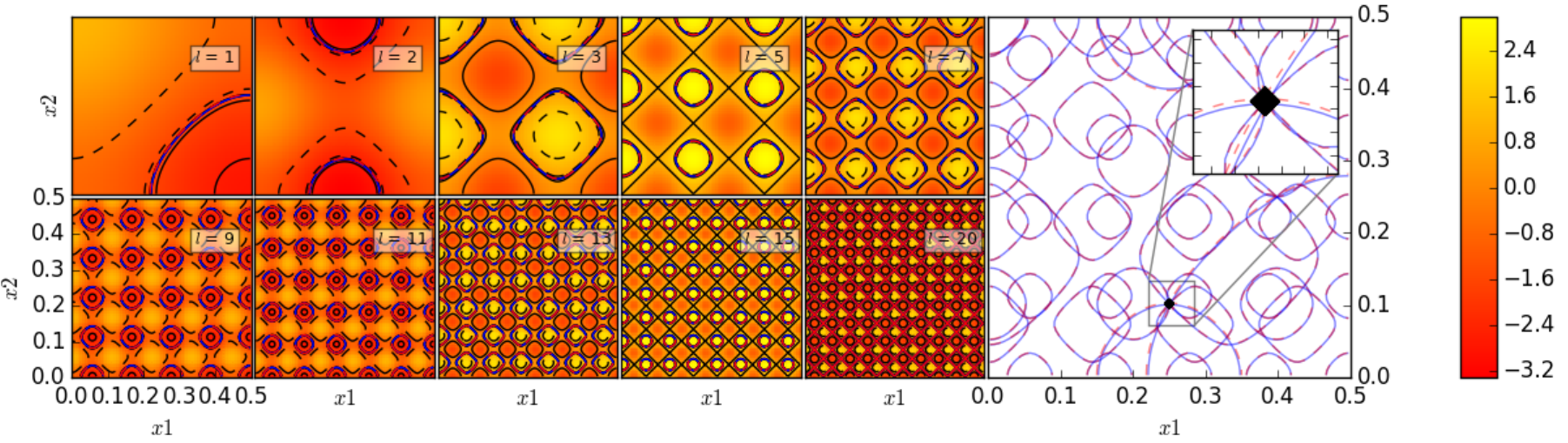}
    \caption{Isosurfaces of geometrical amplitudes for a 1-dim. structure with positions $x_1=0.25, x_2=0.1$ in dependence of form factor sign combinations $(f_1,f_2) = (1,1)$ and $(1,-1)$(top to bottom).}
    \label{fig:structure1}
\end{figure}

In neutron scattering, negative form factors present a specific challenge. This can be illustrated for an easy 1-dim. centrosymmetric structure example of two atoms with equal scattering strength but permutation of sign, \ie $m=2$ and with the respective pairs of form factors $(f_1,f_2) = (1,1)$ and $(1,-1)$. Considering a hypothetical 1-dim. structure with atoms at $x_1 = 0.25, x_2 = 0.1$, the isosurfaces of reflections $l$ and their intersections will give solutions as identified in Fig.~\ref{fig:structure1} with black dot (magnification shown on the upper right).

It has to be noted that the general loss of permutation symmetry by introducing different signs (\eg due to isotope distributions) can in principle be recovered (evident from the shifted mirrors and anti-mirrors in Fig.~\ref{fig:structure1}). The sign change manifests in a shift of the whole parameter space by half a period of the given reflection order along the respective negative coordinate, and thus a relocation of maximum and minimum amplitude centers. This does not alter the general symmetry features of the isosurfaces. If the number of negative signs is known, then these relocations can be predicted, as long as a specific convention of sign order is defined, \eg, first all positive then all negative signs.

The same is true for negative signs of form factors mixed with varying scattering strength, as now partial permutations among atoms with equivalent absolute scattering strength can be identified (in analogy again with predictable shifts of the parameter space along the coordinates with negative sign). As is the case for purely positive signs, high symmetry features are altered, and curvatures (derivatives) of isosurfaces along coordinates of weak scatterers are reduced. Thus, higher correlations to other coordinates may appear, but the PSC formalism is likewise applicable. The generalization to an arbitrary dimension $m$ is also expected since the phase shift introduced by the negative sign only applies to the coordinates of the specific atom, independent of the scattering contributions of all other atoms. 

\subsection{\label{sec:linearization}Linear approximation}
The recent implementation of linearization, based on 1-dim. centric projections of crystal structures elegantly treats structures by mapping experimental or theoretically calculated observations to respective $\isog(h)$ or $\isof(h)$. Fixing the extend and volume of the solution space $\mathcal{S}$ ($\exists\, \mathcal{S} \subseteq \P{m} \mid \partial\isog \subseteq \mathcal{S} \lor \partial\isof \subseteq \mathcal{S} $; subset $\mathcal{S}$ of space $\P{m}$ contains the reduced linear boundaries $\partial\isog(h)$ of $\isog(h)$ or $\partial\isof(h)$ of $\isof(h)$ associated with $h$ reflections in $m$-dim. PS) by the observables, the developed routines, described in Sec. \ref{sec:optionB}, and the potential and performance of PSC linearization have been explored via resolving various synthetic structures in both EPA and non-EPA frameworks by random Monte-Carlo structural parameter distribution, different synthetic structures $\ta{s}^\text{syn}$ generated and resolved. 

\begin{figure}
    \centering
    \includegraphics[width=1\textwidth, clip, trim=2.92in 1.95in 0.14in 1.94in]{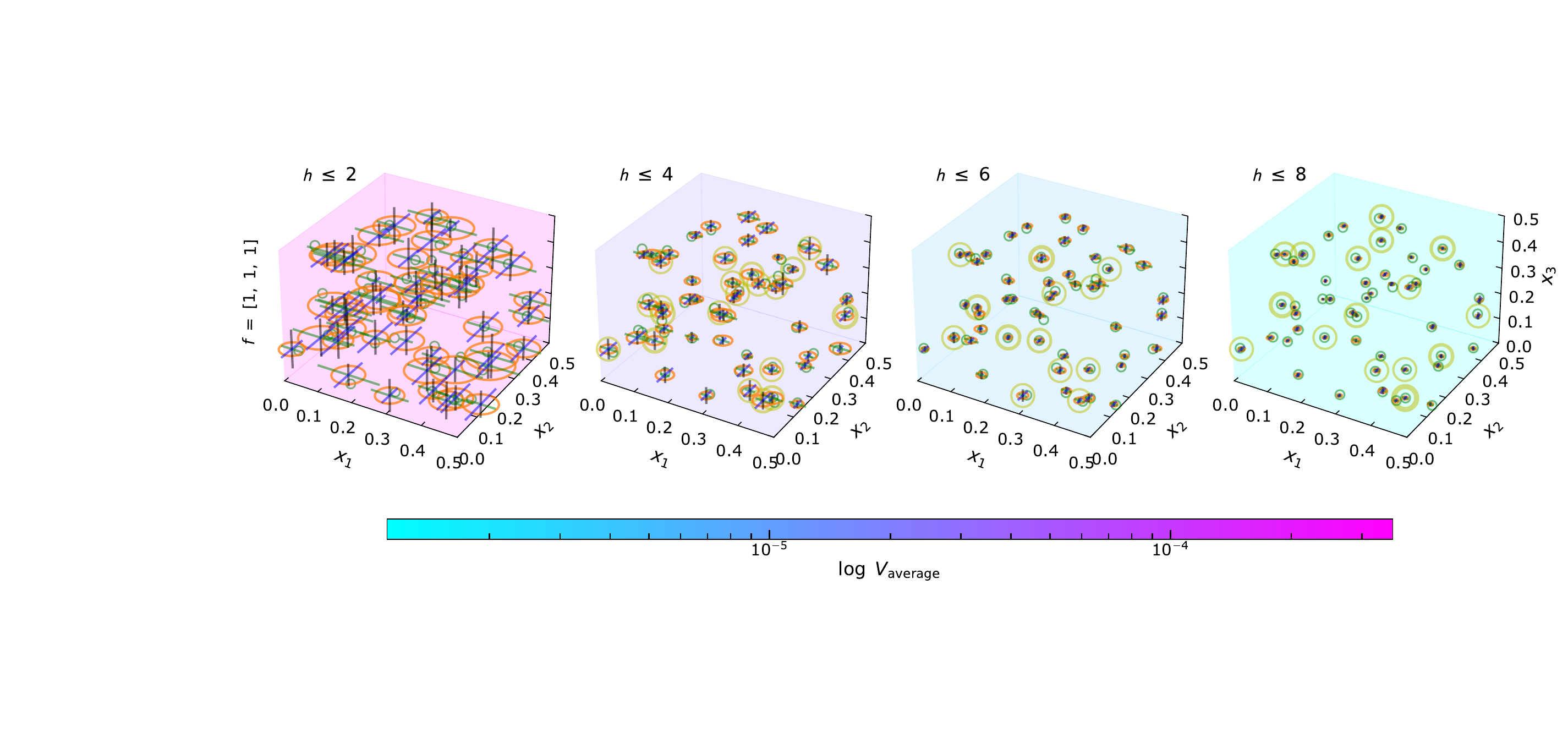}
    \includegraphics[width=1\textwidth, clip, trim=2.92in 1.95in 0.14in 1.94in]{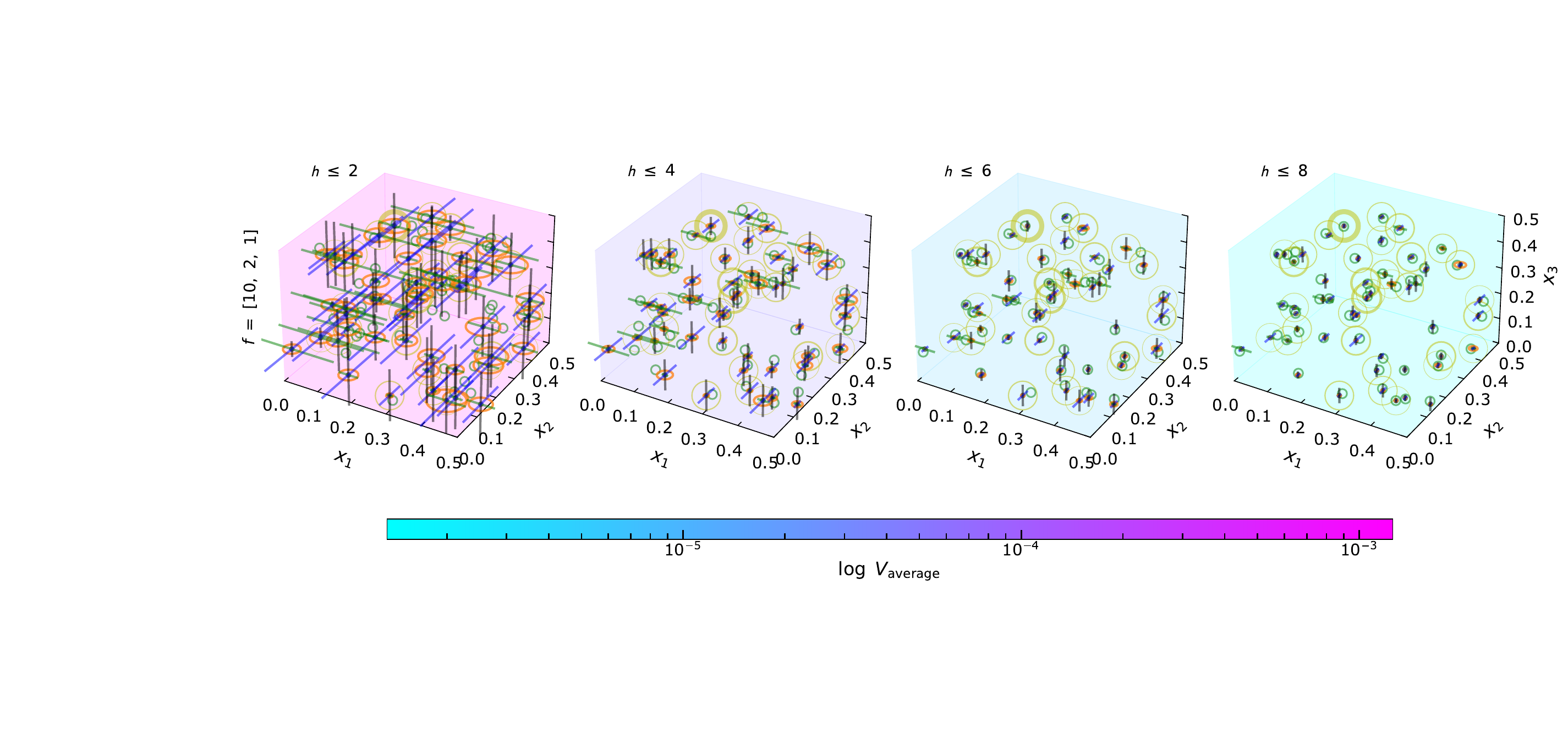}  
    \caption{The resolved synthetic structures $\ta{s}^\text{syn}$ consist of $m=3$ scatterers. The atomic scattering factor combinations $f=[1, 1, 1]$ (top row) and $f=[10, 2, 1]$ (bottom row) theoretically represent equi-point scatterers and heavy-light-light atomic combinations, respectively. The amplitudes of reflections $h$, with $h \leq 2, 4, 6, 8$ are considered within the EPA framework for $f=[1, 1, 1]$, and within the non-EPA framework for $f=[10, 2, 1]$. The green circle represents the centroid of the solution space $\mathcal{S}$ that encircles the $\ta{s}^\text{syn}$. The crossbar on each point indicates the error along each $x_i$ direction, with the number of found solutions represented by the yellow circle, whose thickness increases as the number of solutions increases. The volume $V$ of solution space or spaces (when more than one solution space is found) is summed up to calculate the radius of the virtual sphere as $\mathcal{R} = \sqrt[3]{\nicefrac{3V}{4\uppi}}$, which is represented by orange circles. The background color of each $\P{}$ represents the average volume of $\mathcal{S}$ of all $\ta{s}^\text{syn}$. Results are copied from Ref.~\cite{VNZ2024} with permission.}
    \label{fig:3DMC_EPA_G}
\end{figure}

The results of the EPA framework shown in Fig.~\ref{fig:3DMC_EPA_G} confirm the expected convergence of the found $\ta{a}$ towards $\ta{s}^\text{syn}$ with increasing number of reflections. Without any exception, the true structure $\ta{s}^\text{syn}$ is confined in $\mathcal{S}$, along with secondary homometric and quasi-homometric solution regions, with degeneracies from unique till numerous, depending on the chosen linearization routines and the individual structure. The obtained distinct solution regions in $\mathcal{S}$ possess extent, providing the uncertainty in the found solution. A higher number of experimental observations will, in general narrow down $\mathcal{S}$ further. In addition to the results of the EPA framework, the $\ta{s}^\text{syn}$ shown in Fig.~\ref{fig:3DMC_EPA_G} (top row) are further treated in non-EPA framework by assigning different atomic scattering factors. The results of the non-EPA framework, shown in Fig.~\ref{fig:3DMC_EPA_G} (bottom row). Again, the $\ta{s}^\text{syn}$ are solved with precision down to pm accuracy on varying the atomic scattering powers. 

\begin{figure}
    \centering
    \includegraphics[width=0.48\textwidth]{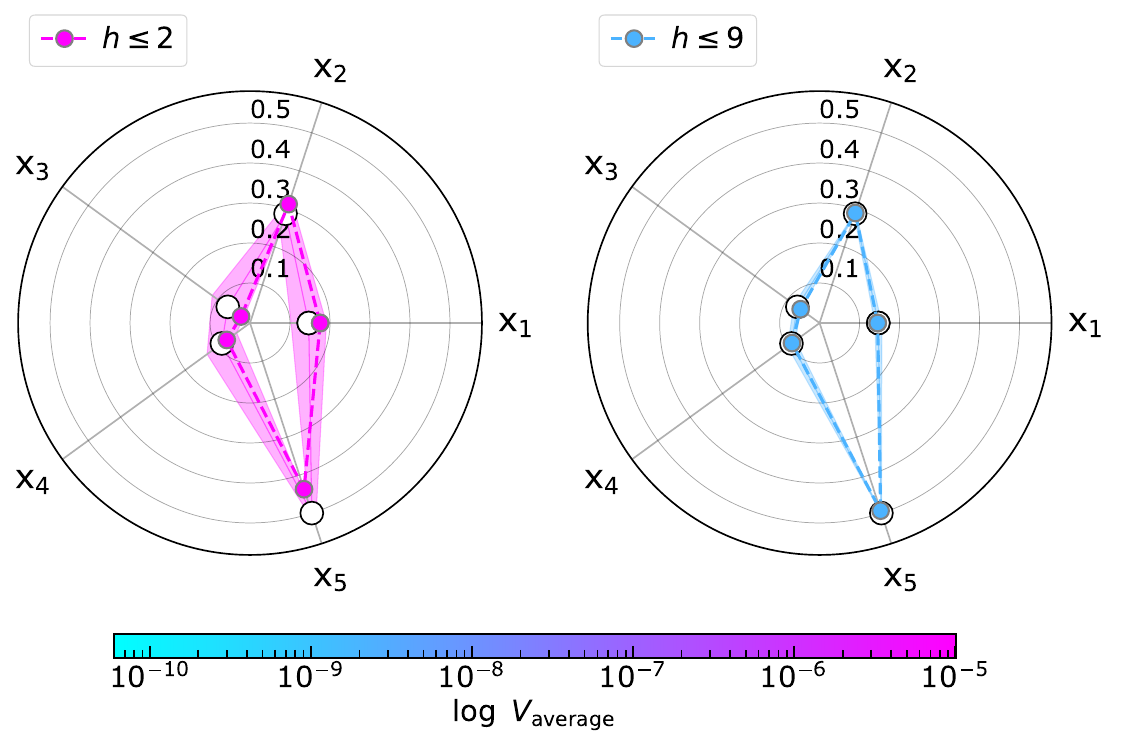}
    \includegraphics[width=0.48\textwidth]{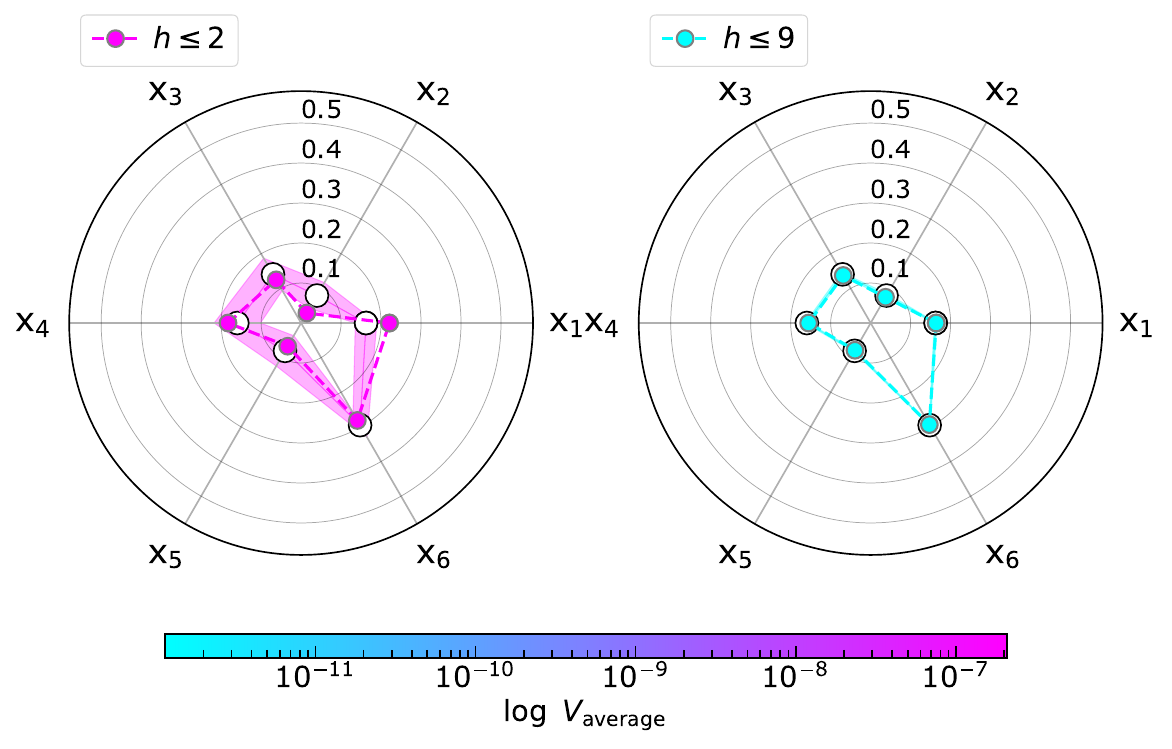}
    \caption{synthetic structure $\ta{s}^\text{syn}$ consisting of five (left) and six (right) atoms, which are treated as equal scatterers. Employing $\P{5}$ and $\P{6}$ and reflections $h$ up to 2 and 9, each $\ta{s}^\text{syn}$ is solved by confining the respective PS to a region of volumes down to \num{1E-9}. The precision as well as accuracy on $\ta{a}$ improves as the amount of reflections $h$ included in the calculation increases. The color-shaded region represents the valid range where each $x_j$ can potentially be positioned. The open and filled circles represent $\ta{s}^\text{syn}$ and $\ta{a}$, respectively}
    \label{fig:5mps}
\end{figure}

The dimension $m$ can be theoretically extended to any required number. Up to date, the algorithms are not yet optimized for performance. They have been tested with structures of five and six atoms and reflections up to $h=9$ in the EPA and non-EPA frameworks, see Fig.~\ref{fig:5mps}.

\section{\label{sec:summary}A consice description on the present knowledge of the PSC}
The novel approaches of PSC for crystal structure determination have been under continuous development over the period of two decades in different phases. Starting from standard scattering principles, models are framed to handle the structures in parameter space, to map reflections as manifolds and to find solution vectors by various strategies. The present knowledge on the PSC-Concept is described shortly in a point-by-point fashion to summarize the key features, advantages, and shortcomings of PSC as follows.

\subsection{Experimental basis} 
Standard Bragg scattering, integrated intensities at points of the reciprocal lattice, perhaps reduced to amplitudes $|F(hkl)|$ or derived ratios present the experimental basis for the PSC approach. 

\subsection{\label{sec:model}Model of the crystal structure}
Two standard descriptions of atomistic crystal structures are used:
% \begin{enumerate}[label=\thesubsection.\arabic*]
\begin{enumerate}[label=(\roman*)]
    \item Atoms as point scatterers (neutron scattering) or (spherical) $f^0 (\sin(\theta/\lambda))\,[+f'(\lambda)\,+\,if^{\prime\prime}(\lambda)]$ for \xrays, both reduced by Debye-Waller thermal smearing.
    \item Structure by coordinates of the $m$ atoms in the asymmetric unit $(x_j,y_j,z_j)$ $(j=1,\ldots,m)$ or $\mathbf{r}_j=x_j\mathbf{a}+y_j\mathbf{b}+z_j\mathbf{c}$ (assumed independent) -- ideally by the sheer geometry, \ie all $f_j$ assumed equal, hence set $f_j=1$, thus $|F(hkl)|$ simplified to the so-called geometrical structure amplitude $g(hkl) = |\sum_{j=1}^{m} e^{2\pi i(hx_j+ky_j+lz_j)}|$.
\end{enumerate}

\subsection{\label{sec:mapping}Mapping of the structure}
Instead of the usual Fourier transform of $F(hkl)$ (scattering density in direct space) or $|F(hkl)|^2$ (interatomic vector density in Patterson space), PSC represents the problem of structure determination as a vector $\mathbf{t} = \mathbf{t}_{x} \otimes \mathbf{t}_{y} \otimes \mathbf{t}_{z}$, with $\mathbf{t}_{x}(x_1, x_2, \ldots, x_m)$, etc. This is done in an orthonormal periodic parameter space of 3$m$-dim., covering all independent positional parameters. In practice, this space forms a finite subspace of the broader Hilbert space. Within $\P{3m}$, we analyze one asymmetric unitcell (see reduction of shortcomings, part (b)).
     
\subsection{\label{sec:solution}Solution of the structure}
Each reflection amplitude (or intensity) $\left|F\left(hkl\right)\right|$ defines a manifold $\mathcal{M}\left(hkl\right)$ of $\le3m-1$ dimensions in $\P{3m}$. Its geometrical representation is called isosurface of (hkl). Thus, in principle, the solution vector $\ta{a}$ is found by the intersection of at least $3m$ independent $\mathcal{M}$. The definition of isosurfaces in the PSC is, to our best knowledge, new in crystallography, but was found helpful for structure determination. 

\subsection{\label{sec:features}Features of PSC}
\begin{enumerate}
    \item\label{item:651}  Instead solving either the phase problem of the amplitudes, or the de-convolution problem of a Patterson map, both from many measured reflection amplitudes $\left|F\left(hkl\right)\right|$ or intensities $\left|F\left(hkl\right)\right|^2$ (or in short: of $\mathbf{h}$ instead of $hkl$), we obtain (for a 1-dim. acentric structure of five atoms plus a sixth defining the origin) from the Patterson (or vector set) Matrix by setting $x_m=0$, see Fig. 3 and Tab. 1 in Ref. \cite{ZF09}, the combined solution \\
     
     \begin{center}
     $\begin{pmatrix}
            &     &     &     &     & -x_1 \\
            &     &     &     &     & -x_2 \\
            &     &     &     &     & -x_3 \\
            &     &     &     &     & -x_4 \\
            &     &     &     &     & -x_5 \\
        x_1 & x_2 & x_3 & x_4 & x_5 & 0
     \end{pmatrix}$ \\[1cm]
     \end{center}
     
     from a few, say here $\ge m$ absolute reflection (amplitudes or intensities, assumed independent). It is also sorted \textit{vs.} rows ($x_j$) and columns ($-x_j$) hence we get the complete structure solution $\ta{s}$ (\ie a good approximation $\mathbf{t}\approx\ta{s}$) separated from its inverse $-\ta{s}$ however not knowing which of the two is the correct one.
     
     \item\label{item:652} The resolution $\Delta$ for atomic positions is much smaller than defined by $(\lambda/\sin(\theta_{\max})$: in principle, $\Delta=0$ for $n \ge 3m$ independent amplitudes assumed free of error.
     
     \item\label{item:653} Should 
     \begin{enumerate}[label=(\roman*)]
         \item just one solution $\ta{s}$ be found, it is unique (for the reflections involved)
         \item more than one solution be found, then all are presented at once.
     \end{enumerate} 
     If all their reliability factors (any $mathrm{FOM}$) are identical, they are called homometric solutions;  slightly different FOM referred to as pseudo-homometric.
     
     \item\label{item:654} $\mathcal{M}$ is in general a piecewise analytic hyper-surface (smoothly differentiable) for structures with known end coordinates at the $3m$ axes of $\mathbf{\P{3m}}$ and at its main diagonal $[1,\, 1,\, \ldots,\,1]$, thus easy to handle, provided it is based on $\left|F\right|$, $\left|F\right|^2$, or simple, related entities as, \eg the unitary structure amplitude $|U(\mathbf{h})|\equiv|F(\mathbf{h})|/\sum_{j=1}^{m}f_j$, or the geometric structure amplitude $g(\mathbf{h})=|\sum_{j=1}^{m}\cos{2\pi l x_j}|$ (for 1-dim. centrosymmetric structures) etc. This is not true for, \eg, $|E(\mathbf{h})|$ of the Direct Methods and for the $e(\mathbf{h})$ and  $q(\mathbf{h},\mathbf{k})$ derived in Sec.~\ref{sec:gorIratio} and Sec.~\ref{sec:quasig}. $\left|F\right|$ or $\left|F\right|^2$ may be used at will. (If partial structures are in play, $\left|F\right|$ are easier to handle, because no cross-products between different partial structure amplitudes exist.)
    
    \item\label{item:655} The $\geq3m$ reflections selected after (i) may be any  random selection and may be applied in any sequence expected to be useful. Thus, a complete data set is not necessary.
    
    \item\label{item:656} Each zero amplitude $\left|F(\mathbf{h})\right| = 0$ acts as a constraint to the unknown positional parameters, thus reducing their effective number by one (centric structures) or by two (acentric ones).
    
    \item\label{item:657} The picture $\mathcal{M}$ in parameter space of any maximal possible $F(\mathbf{h}) = |\sum_{j=1}^{m}{{f_j}(\mathbf{h})}|$ is a finite number of points (depending on $\mathbf{h}$) at positions known \textit{a priori}, instead of a continuous isosurface. $(m-1)$-dim. nearly hyper-spherical  (\ie in dimensions $>$3) approximations for isosurfaces around these points are useful for $F(\mathbf{h})\approx|\sum_{j=1}^{m}{f_j(\mathbf{h})}|$. A trivial consequence of this statement and the last (\ref{item:656}) is: high contrast in a few amplitudes is more helpful towards a solution than many amplitudes of average size.
    
    \item\label{item:658} Computationally demanding when compared to, \eg, Direct Methods. Its order of magnitude is proportional to $n_\mathrm{step}^{3m}$, see Sec.~\ref{sec:preference}, for a wanted $\Delta<1/n_\mathrm{step}$ (applying Option A, see Sec.~\ref{sec:optionA}) to estimate what can be reached (known as curse of dimensions or combinatorial explosion). The consequence is that a soluble $m$ is limited for all computational developments.
    
    \item\label{item:659} The method and thus its results are vulnerable to errors: In conventional methods, even a substantial error of a single reflection amplitude is compensated by the high and average quality amplitudes. In PSC, it may become fatal because its $\mathcal{M}$ leads into a false region of Parameter Space. The absolute scale factor of all amplitudes is a special problem.
    
    \item\label{item:6510} The equal point atom model, if approximated, has limits hitherto unknown. A similar problem applies to sign differences of $f_j$ in neutron diffraction \cite{panet01217} (\eg, for H and D nuclei).
    
    \item\label{item:6511} Advances (i) and (v) may hinder an easy development of a black-box system for avoiding intelligent interrupt(s) during the solution process.
\end{enumerate}

\subsection{\label{sec:reduction}Reduction of some shortcomings}
\subsubsection{\label{sec:a}Limiting to one-dimensional problems} \ie, to 1-dim. projections of a 3-dim. or 2-dim. structure, \eg to $\ta{x}$, results in elegant advantages over traditional methods while imposing unavoidable  disadvantages.

\noindent\text{Advantages}
\begin{enumerate}[label=(a\arabic*)]
    \item $\ta{s,x}$ is based on less reflections, \eg on $|F(h)|$
    \item drastic reduction of computing demand ($n_s^{3m} \rightarrow n_s^m$)
    \item re-construction of 3-dim. structure is possible from $\geq5$ 1-dim. projections, also via 2-dim. interim projections.
\end{enumerate}

\noindent\text{Disadvantages}
\begin{enumerate}[label=(d\arabic*)]
    \item much higher probability for homometric and pseudo-homometric solutions
    \item correlation effects by exact or nearly exact overlapping of atoms in the projection
    \item chemical constraints, \eg bond distances, binding angles cannot be applied
    \item re-construction is also principally limited by a finite $m$ (this limit has been not reached, however, until now).
\end{enumerate}

\subsubsection{\label{sec:b}Equal-Point Atom model}
The EPA model, applied to $m$ atoms, fosters a permutation symmetry of order $m!$, \eg a computational reduction of $>10^6$ for $m=10$, since only the asymmetric part $\A{m}$ of $\P{m}$ must be searched.

\noindent\text{Advantages}
\begin{enumerate}[label=(a\arabic*)]
    \item drastic reduction of computing demand
\end{enumerate}

\noindent\text{Disadvantages}
\begin{enumerate}[label=(d\arabic*)]
    \item if the EPA model is applied only to a subset of $m^\prime$ atoms with $m^\prime \le m$, a smaller permutation symmetry results of order $m^\prime !$
    \item see feature~\ref{item:6510} about EPA limitations in Sec.~\ref{sec:features}
\end{enumerate}

\subsubsection{\label{sec:c}Absolute scale factor}
The absolute scale factor on the $|F_{obs}|$ can be replaced by quotients of amplitudes $q(\mathbf{h},\mathbf{k})$ or by normalization to average $e(\mathbf{h})$.

\noindent\text{Disadvantages}
\begin{enumerate}[label=(d\arabic*)]
    \item see feature~\ref{item:654} about scaling restrictions in Sec.~\ref{sec:features}
\end{enumerate}

\subsection{\label{sec:What}What can be achieved in practice}
\subsubsection{\label{sec:data}Data} 
By using modern synchrotron radiation facilities, reproducibility of reflection intensities is presently about 1\,\% (sometimes better). Hence, the importance of the old rule reflection phases are more important than amplitudes is thus diminished for structure solution.

\subsubsection{\label{sec:methods}Methods} 
Up to now, 3 options (see Sec.~\ref{sec:optionA},~\ref{sec:optionB}, and~\ref{sec:optionC}) have been found how to evaluate $\mathbf{t}$ by PSC-techniques from, say, reflection amplitudes $|F|$:
\begin{enumerate}
    \item Option A (see Sec.~\ref{sec:optionA}) calculates any kind of Figure of Merits FOM for the grid-points in $\A{m}$ and finds the global best $\ta{s}$ assuming $\ta{s} - \mathbf{t} \approx 0$. Thus, no isosurfaces are needed.
   
    \item Option B (see Sec.~\ref{sec:optionB}) derives the intersection point of all isosurfaces $\mathcal{M}(\mathbf{h})$, leads in principle to one (or more) $\ta{s}$, meeting the corresponding experimental results.
    
    \item In Option C (see Sec.~\ref{sec:optionC}), $\A{m}$ is reduced step-wise by inequalities between selected pairs of $\mathcal{M}(\mathbf{h}_1)$ and $\mathcal{M}\left(\mathbf{h}_2\right)$ isosurfaces, followed by more such pairs, until the resulting $m$-dim. solution region(s) $\mathcal{S}$ is/are small enough. Thus, one or more $\ta{s}$ is/are found, each \eg as its center of gravity.
\end{enumerate}

For Option B and C, a numerical description or approximation of $\mathcal{M}$ is mandatory, hence Option C (in particular) has a more academic character.

In principle, also non-geometric parameters, \ie parameters not reflecting atomistic positions in the crystallographic unit cell (individual atomic anisotropic displacement parameters, scaling factor, etc.) affecting  
$|F_\mathrm{obs}(\mathbf{h})|$ can be introduced into $\mathbf{P^m}$ as additional dimensions of the structure vector $\mathbf{t}$.

% \subsubsection{\label{sec:parameter}Parameters} 

\subsection{\label{sec:whathasbeen}What has been reached in practice}
Using Option A at a standard PC, centric structures with $m \le 11$ has been solved within hours. For heavy-atom structures, it could be achieved~\cite{KF10}: precise locations of (a few) heavy atoms were found together with a precise scale factor (as an additional PSC parameter). Options B and C have been found working well for a few examples having small $m$. 

\subsection{\label{sec:whatmustbe}What must be done soon?}
There is a lack of theoretical background for the application of PSC to structure amplitude equations and their corresponding isosurfaces, in particular with regard to acentric structures and verification aspects: data quality and quality of results.

\subsection{\label{sec:Generalaims}General aims to consider}
If the complete structure is unknown (and conventional methods did not succeed), PSC can be used as described. If, however, only a part of the structure is needed and the coordinates of all other atoms are known, two helpful consequences may arise:
\begin{enumerate}
    \item The signs of some reflection amplitudes are safely known (or the phases approximately).
    \item If the unknown atomistic positions are on special Wyckoff equipoints, perhaps only a reduced number of data must (or can) be applied.
\end{enumerate}
Advantages of PSC methods over standard techniques (\eg, high spatial resolution, unique or all possible solutions) may be considered thoroughly.

\subsection{\label{sec:whatdowenotwant}What do we not want}
We would feel unreasonable if trying to compete with conventional methods as, \eg, of type SHELX, which solve structures in a black box routine.

\section{\label{sec:conclusion}Conclusion and Outlook}
The Parameter Space Concept, a structure determination technique proposed and developed by Fischer \etal over the last decades, has been successfully implemented and realized through current advancements in computational resources. In recent years, a concrete workflow has been developed to handle theoretical observations and to solve crystal structures using different methods, such as grid-based least-squares fitting and linearization of scattering amplitudes, within PSC. 

The applied concept of 1-dim. projections for 3-dim. crystal structures simplifies the computational demand of the structure determination from diffraction data. In the core of the linearization techniques, the scattering amplitudes or intensities are interpreted as piecewise analytic hypersurfaces in the parameter space, and a common intersection point or region of these hypersurfaces of different observations reveals the crystal structures of interest. 

Offering options to approximate all scatterers as having the same scattering power or to treat them with energy-dependent scattering power, PSC handles both approximated geometric and realistic structures. By those means, the challenge of crystal structure determination has been simplified into a geometric problem in an $m$-dim. parameter space. With a limited number of reflections, a solution space with accuracy in the range of \si{\pico\meter} may be determined by PSC in the $m$-dim. parameter space. A further advantage of PSC is that, along with the homometric solutions, potentially equally probable non-homometric structures in accordance with the experimental diffraction data are also determined in a single analysis. Exclusively for the linearization techniques, a Python-based code has been developed and made publicly available on GitHub~\cite{code}.

Though capable of resolving real crystal structures, PSC suffers from different factors and opens up new venues for further improvement. The present impediments originate from: (i) \textbf{Symmetry constraints}: the PSC implementations do not include crystal structure symmetry beyond $\texttt{P1}$ and $\texttt{P}\bar{1}$, all structures are treated in those low symmetry groups. Specifically for the linearization routines, solvable structures are still limited to centrosymmetry. \textbf{(ii) Data handling}: the linear equations to reduce the complete PS to solution space grow with the rate of $2\cdot n^m$ ($n$ number of considered reflections). The experience of data handling in the current status indicates that efficient data handling as well as parallelization of the currently coded routines play a crucial role in the generalization of the PSC method to $m$-dim. PSs with $m > 10$. \textbf{(iii) Reduction of non-homometric solutions}: the error envelopes induced by linear approximation may result in false-minima solutions distinctively away from the true structure, requiring further reduction of such solutions (\eg, by an increased number of anchor points in the linearization process). \textbf{(iv) Reduction of computational time}: processing the experimental or theoretical amplitudes using any methods in PSC (grid-based approximation or linearization) demands huge computational resources with the current implemented techniques. For example, filling the complete PS with linear boundaries of amplitudes and set operations are time-consuming processes. Therefore, optimization of those processes is inevitable for PSC application in higher-dimensional PS, such as for $m>10$. \textbf{(v) Selection of reflection sequence}: Obtaining a small $\ta{s}$ at the earliest stage of structure determination is beneficial to reduce the computational cost, which grows significantly with increasing parameter space dimension $m$. A systematic study which Evaluation Order provides the most efficient entry into the PSC process is yet to be carried out.

Since only a few experimental observations are sufficient, the PSC is especially interesting for studying time-dependent processes in already-known crystal structures, \eg, during phase transformations, dynamic transport processes, etc. In most cases, these reflections can be precisely specified in a preliminary analysis. This makes the PSC very attractive for \textit{in-situ} diffraction at synchrotron beamline to localize or detect the smallest changes in atomic positions in material systems with known atomistic models, which meets the current demands of \textit{in-situ} and \textit{in-operando}.

\section{\label{sec:symbols}Symbols and Definition}

\begin{table}
    \centering
    \caption{Glossary: List of symbols and definitions used in PSC.}
    \resizebox{0.6\textwidth}{!}{
    \begin{tabular}{c|c}
        
        \hline\hline
        Symbol & Definition \\
        \hline\hline

        $f$                 &   atomic scattering factor    \\
        $F_{\mathrm{obs}}$  &   observed structure factor \\
        $(hkl)$             &   reflection, denoted with Miller indices \\
        $j$                 &   index of atom in unit cell \\
        $K$                 &   scaling factor \\
        $m$                 &   number of atoms in unit cell \\
        $\P{3m}$            &   $3m$-dimensional parameter space \\
        $\bf{r}$            &   vector of fractional coordinates of an atom \\
        $\bf{t}$            &   structure vector \\
        $\bf{t}_a$          &   vector of the approximate structure solution \\
        $\bf{t}_s$          &   true structure vector \\
        $\bf{t}_t$          &   test vector \\
        $V$                 &   unit cell volume \\
        $\rho(\bf{r})$      &   scattering density of an atom at position $\bf{r}$ \\
        $\Delta t$          &   equal grid spacing for all dimensions \\

        \hline\hline
    \end{tabular}
    }
    \label{tab:glossary}
\end{table}

\begin{acknowledgments}
This article is dedicated to the enduring legacy of the respected Prof. Dr. Karl Fischer (deceased in 2023), whose profound contributions to crystallography continue to inspire us. He was an invaluable part of the PSC project, and we shall remember the valuable lessons learned under his insightful guidance and unwavering passion for knowledge. Prof. Fischer actively participated in weekly video conferences from his retirement home until shortly before the end of his life and already sketched out a general outline of this review. His intellectual freshness, professional and expressly human advice, and constant joy in exchanging ideas, even with young representatives of the next generation, were highly enriching for this work.

Special thanks to the Blablador tool for providing valuable assistance in structuring and enhancing this article. Its innovative features significantly improved the efficiency and quality of our writing process.

MZ, MV, MN, and DCM acknowledge funding by the DFG within the project DFG 442646446, ZS 120/51.
\end{acknowledgments}

\bibliography{iucr}
\end{document}